\documentclass[useAMS,a4,usenatbib]{mn2e}
\input psfig.sty

\usepackage{floatflt, epsfig}

\def \chisq  {\ifmmode  \chi^2   \else  $\chi^2$  \fi}  
\def \spose#1{\hbox  to 0pt{#1\hss}}  
\def \lta{\mathrel{\spose{\lower 3pt\hbox{$\sim$}}\raise  2.0pt\hbox{$<$}}}
\def \gta{\mathrel{\spose{\lower  3pt\hbox{$\sim$}}\raise 2.0pt\hbox{$>$}}}

\def \kms {\ifmmode  \,\rm km\,s^{-1} \else $\,\rm km\,s^{-1}  $ \fi }
\def \kpc {\ifmmode  {\rm~kpc}  \else ${\rm~kpc}$\fi}  
\def \pc {\ifmmode  {\rm~pc}  \else ${\rm~pc}$ \fi  }  
\def \Gyr {\ifmmode  {\rm~Gyr}  \else ${\rm~Gyr}$\fi}
\def \Msun {\ifmmode M_{\odot} \else $M_{\odot}$ \fi} 
\def \Lsun {\ifmmode L_{\odot} \else $L_{\odot}$ \fi} 
\def \Rsun {\ifmmode R_{\odot} \else $R_{\odot}$ \fi} 
\def \Msunpyr {\ifmmode M_{\odot}{\rm~yr}^{-1} \else $M_{\odot}{\rm~yr}^{-1}$ \fi} 
\def \hMsun {\ifmmode h^{-1}\,\rm M_{\odot} \else $h^{-1}\,\rm M_{\odot}$ \fi}

\def \LCDM {\ifmmode \Lambda{\rm CDM} \else $\Lambda{\rm CDM}$ \fi}
\def \sig8 {\ifmmode \sigma_8 \else $\sigma_8$ \fi} 
\def \OmegaM {\ifmmode \Omega_{\rm M} \else $\Omega_{\rm M}$ \fi} 
\def \OmegaL {\ifmmode \Omega_{\rm \Lambda} \else $\Omega_{\rm \Lambda}$\fi} 
\def \Deltavir {\ifmmode \Delta_{\rm vir} \else $\Delta_{\rm vir}$ \fi}
\def \rhocrit {\ifmmode \rho_{\rm crit} \else $\rho_{\rm crit}$ \fi}
\def \rhou {\ifmmode \rho_{\rm u} \else $\rho_{\rm u}$ \fi}
\def \zc {\ifmmode z_{\rm c} \else $z_{\rm c}$ \fi}

\def \rhos {\ifmmode \rho_{\rm s} \else $\rho_{\rm s}$ \fi} 
\def \rs {\ifmmode r_{\rm s} \else $r_{\rm s}$ \fi} 
\def \cvir {\ifmmode c_{\rm vir} \else $c_{\rm vir}$ \fi} 
\def \Rvir {\ifmmode r_{\rm vir} \else $R_{\rm vir}$ \fi}
\def \Vvir {\ifmmode V_{\rm  vir} \else  $V_{\rm vir}$  \fi} 
\def \Mvir {\ifmmode M_{\rm  vir} \else $M_{\rm  vir}$ \fi}  
\def \Nvir {\ifmmode N_{\rm  vir} \else $N_{\rm  vir}$ \fi}  
\def \Jvir {\ifmmode J_{\rm vir} \else $J_{\rm vir}$ \fi} 
\def \Evir {\ifmmode E_{\rm vir} \else $E_{\rm vir}$ \fi} 
\def \vvir {\ifmmode v_{\rm vir} \else $v_{\rm vir}$ \fi} 
\def \lam {\ifmmode \lambda  \else $\lambda$ \fi} 
\def \lamp {\ifmmode \lambda^{\prime} \else $\lambda^{\prime}$  \fi} 
\def \Vmax {\ifmmode V_{\rm  max} \else  $V_{\rm max}$  \fi} 
\def \Mdm {\ifmmode M_{\rm  dm} \else $M_{\rm  dm}$ \fi}

\def \Mgas {\ifmmode M_{\rm gas} \else $M_{\rm gas}$ \fi} 
\def \Mcg {\ifmmode M_{\rm cg} \else $M_{\rm cg}$\fi} 
\def \Mhg {\ifmmode M_{\rm hg} \else $M_{\rm hg}$ \fi} 
\def \Mdisc {\ifmmode M_{\rm disc} \else $M_{\rm disc}$ \fi} 
\def \Md {\ifmmode M_{\rm d} \else $M_{\rm d}$ \fi} 
\def \Mda {\ifmmode M_{\rm d,0\%} \else $M_{\rm d,0\%}$ \fi} 
\def \Mdb {\ifmmode M_{\rm d,20\%} \else $M_{\rm d,20\%}$ \fi} 
\def \Mdc {\ifmmode M_{\rm d,40\%} \else $M_{\rm d,40\%}$ \fi} 
\def \md {\ifmmode m_{\rm d} \else $m_{\rm d}$ \fi} 
\def \Mb {\ifmmode M_{\rm b} \else $M_{\rm b}$ \fi} 
\def \Mbh {\ifmmode M_{\rm b,pri} \else $M_{\rm b,pri}$ \fi} 
\def \Mbs {\ifmmode M_{\rm b,sat} \else $M_{\rm b,sat}$ \fi} 
\def \zo {\ifmmode z_{0} \else $z_{0}$ \fi} 
\def \rd {\ifmmode r_{\rm d} \else $r_{\rm d}$ \fi}
\def \rg {\ifmmode r_{\rm g} \else $r_{\rm g}$ \fi}
\def \rb {\ifmmode r_{\rm b} \else $r_{\rm b}$\fi}
\def \rs {\ifmmode r_{\rm s} \else $r_{\rm s}$\fi}
\def \rc {\ifmmode r_{\rm c} \else $r_{\rm c}$\fi}
\def \rvir {\ifmmode r_{\rm vir} \else $r_{\rm vir}$\fi}
\def \rbh {\ifmmode r_{\rm b,pri} \else $r_{\rm b,pri}$ \fi} 
\def \rbs {\ifmmode r_{\rm b,sat} \else $r_{\rm b,sat}$ \fi}

\title[Simulated galaxy merger trees] 
{Numerical hydrodynamic simulations based on semi-analytic galaxy merger trees: method and Milky-Way like galaxies}

\author[B. P. Moster et al.] {Benjamin P. Moster$^{1}$
 \thanks{moster@mpa-garching.mpg.de}, Andrea V. Macci\`o$^{2}$, Rachel S. Somerville$^{3}$\\ 
  $^1$ Max-Planck Institut f\"ur Astrophysik, Karl-Schwarzschild Stra\ss e 1, 85748 Garching, Germany\\
  $^2$ Max-Planck-Institut f\"ur Astronomie, K\"onigstuhl 17, 69117 Heidelberg, Germany\\ 
  $^3$ Department of Physics and Astronomy, Rutgers University, 136 Frelinghuysen Rd., Piscataway, NJ 08854
}

\begin{document} 
              
\date{\today}
              
\pagerange{\pageref{firstpage}--\pageref{lastpage}}\pubyear{2011} 
 
\maketitle 

\label{firstpage}
             
\begin{abstract}
  
We present a new approach to study galaxy evolution in a
cosmological context.  We combine cosmological merger trees and
semi-analytic models of galaxy formation to provide the initial
conditions for multi-merger hydrodynamic simulations. In this way we
exploit the advantages of merger simulations (high resolution and
inclusion of the gas physics) and semi-analytic models
(cosmological background and low computational cost), and integrate
them to create a novel tool.
This approach allows us to study the evolution of various galaxy
properties, including the treatment of
the hot gaseous halo from which gas cools and accretes onto the
central disc, which has been neglected in many previous studies.
This method shows several advantages over other methods. As only the
particles in the regions of interest are included, the
run time is much shorter than in traditional cosmological simulations,
leading to greater computational efficiency. 
Using cosmological simulations, we show that multiple mergers are
expected to be more common than sequences of isolated mergers, and
therefore studies of galaxy mergers should take this into account. 
In this pilot study, we present our method and illustrate the results
of simulating ten Milky Way-like galaxies since $z=1$. We find good
agreement with observations for the total stellar masses, star
formation rates, cold gas fractions and disc scale length parameters.
We expect that this novel numerical approach will be very useful for
pursuing a number of questions pertaining to the transformation of
galaxy internal structure through cosmic time.

\end{abstract}


\setcounter{footnote}{1}

\section{Introduction}
\label{sec:intro}

Large galaxy redshift surveys, such as the Two Degree Field Galaxy
Redshift Survey \citep{folkes1999} and the Sloan Digital Sky Survey
\citep{stoughton2002} have measured and quantified the large scale
structure of the Universe. On the other side, computational advances
also allow accurate predictions of the distribution of dark matter on
large scales based on numerical simulations within the $\Lambda$ Cold
Dark Matter (CDM) paradigm. The predictions of $\Lambda$CDM are in
excellent agreement with observations on large scales.

On small scales however, it is much more difficult to explain and
predict galaxy structures as this requires an understanding of the
baryonic processes that shape galaxy properties.  The standard picture
of hierarchical galaxy formation was introduced by \citet{white1978}
and has been subsequently extended. In this picture, small initial
density fluctuations grow with time due to gravitational
instability. While the dark matter collapses into haloes with a
quasi-equilibrium state through violent relaxation, the baryonic
matter falls into the potential wells of these haloes forming a hot
gaseous halo for which the self-gravity is balanced by pressure
gradients. The gas in this halo is able to cool, which reduces the
pressure support and causes the accretion of cold gas from the halo
onto a central disc. Here, stars can form depending on the gas density
on a characteristic time-scale resulting in a rotating stellar disc. A
fraction of the newly created stars is short-lived and these stars
explode in supernovae (SNe) which can heat the surrounding gas,
reducing the efficiency of star formation and in some cases blowing
gas out of the galaxy in a galactic wind. Another potential source of
feedback is provided by the accretion of gas onto a supermassive black
hole (BH) in the centre of the galaxy, called active galactic nuclei
(AGN).

Due to the hierarchical nature of the $\Lambda$CDM scenario, a dark
matter halo constantly accretes new material and eventually other
galactic systems merge with it. Such a merger may be accompanied by a
strong burst of star formation if the merging galaxies have a similar
mass and contained significant amounts of cold gas. At the same time,
angular momentum is transferred to the stars in the disc.  For a major
merger, the orbits of the disc stars are randomised, resulting in the
destruction of the discs and the creation of an elliptical
galaxy. After such a merger a new gas disc can be created and a new
stellar disc formed \citep{moster2011a}.  In minor mergers, the disc
of the central galaxy typically survives, although it may be thickened
and some of the stars can be removed from the disc and transferred to
a spherical component while the satellite is completely destroyed
\citep{moster2010b,moster2011b}.  In this way, galaxy mergers affect
the morphology of galaxies and shape their properties.

One of the major goals of galaxy formation modeling is to understand
the correlations of galaxy internal properties with their formation
history and environment. In addition to global properties such as
luminosity or stellar mass, star formation rate, and color, we can
also study the internal structure and kinematics of galaxies.  For example,
there are well-known scaling relations between galaxy luminosity or
stellar mass, radial size, and rotation velocity or velocity
dispersion (called the Fundamental Plane; the Tully-Fisher relation
for late type galaxies or Faber-Jackson for early types are
projections of this plane). Moreover, there are strong correlations
between galaxy morphological or structural properties (e.g. spheroid
vs. disk dominated) and spectrophotometric properties (color or
specific star formation rate), which are not yet fully understood. 

In order to study the connection between large scale environment and
galaxy internal properties, we can specify several requirements for a
galaxy formation modeling approach. The first prerequisite is the
inclusion of gas and the physical processes to which it is
subject. While this may seem obvious, there are many studies in the
literature that neglect the gaseous component and use initial
conditions that include only the collisionless components of stars and
dark matter. However, not only does the gas form stars, subsequently
altering the density and gravitational potential, but the gas is also
able to radiate energy away through cooling. By this mechanism, the
total energy content of a system can be reduced, which is not possible
in pure collisionless studies. The next requirement is high enough
resolution to capture the internal structure of galaxies. Many
cosmological simulations of galaxy formation have $\sim$ kpc spatial
resolution, which clearly does not allow us to say anything about the
internal structure or morphology of galaxies. Third, it is important
to study the evolution of galaxies within a cosmological
background. The cosmological framework has a large impact on the
evolution of a galaxy. It specifies when a halo and a galaxy form and
how much mass they accrete during their evolution. Furthermore, the
cosmology determines the merger histories of the galaxies, i.e. when
and how often a galaxy undergoes a merger event. Finally, while it can
be useful to study a small number of systems in great detail, in order
interpret the significance of any results and make use of the
available large observational samples of galaxies, it is helpful to be
able to simulate a statistically significant sample of systems,
spanning a range of the relevant properties such as halo mass,
formation history, etc. In this paper we present a novel approach that
allows us to carry out numerical hydrodynamic simulations of galaxies
at high resolution, within a proper cosmological context, with much
greater computational efficiency than standard methods.

\subsection{Numerical simulations}

Numerical simulations are a powerful tool for studying galaxy
formation and evolution. Dissipationless $N$-body methods, which
neglect the gas physics, have been extensively employed.
State-of-the-art $N$-body simulations include the \lq
Millennium\rq~simulation \citep{springel2005d}, the \lq Bolshoi\rq~simulation
\citep{klypin2011}, the \lq Via Lactea\rq~simulation
\citep{diemand2007}, the \lq Aquarius\rq~project \citep{springel2008}
and the \lq GHALO\rq~simulations \citep{stadel2009}.
Collectively, N-body simulations have characterized the formation of
dark matter structures from the internal structure of dark matter
halos, to the largest structures we expect to form on hundreds of Mpc.

When a dissipational component is included in the simulation,
the complexity of the problem increases substantially as does the
computational time. One consequence is that cosmological
hydrodynamical simulations typically cannot reach the same spatial and
mass resolution of $N$-body runs. Due to the limited resolution, as
well as imperfect treatment of feedback processes, cosmological
hydrodynamic simulations have long suffered from a set of intertwined
problems. They have tended to produce galaxies that are too compact
and too bulge-dominated (the angular momentum catastrophe), and with
stellar or baryonic masses that are too large compared with their halo
masses (the overcooling problem). They do not typically reproduce the
red colors and very low star formation rates of observed massive
galaxies (another manifestation of the overcooling problem). 
Recent efforts have demonstrated that with very high resolution,
achievable through ``zoom'' techniques, as well as improved treatment
of star formation and stellar feedback, it may be possible to produce
disk-dominated galaxies with sizes and baryon fractions that are
consistent with observations. \citep{governato2004,robertson2004,
okamoto2005,governato2010,agertz2011,guedes2011,stinson2012}. 
This is encouraging, but a drawback of this approach is the enormous
computational expense. Even with very large expenditures of
computational resources, it is only feasible to simulate a handful of
systems at this resolution, particularly if one wishes to run the
simulations to $z=0$.

A third approach neglects the cosmological background and simulates an
isolated galaxy or the interaction of two pre-formed galaxies in a
binary merger. With this technique, it is possible to study galaxy
transformations at very high resolution employing gas physics
\citep{noguchi1988,
  combes1990,mihos1996,cox2006b,naab2006,robertson2006,moster2011a}.  
However, in this approach, the initial conditions (such as the
properties of the progenitor galaxies and the orbit) are not derived
from a cosmological context, but are specified a priori, based on
observations or simply spanning a desired range of values. Perhaps
more importantly, on the timescale of the merger process (a few Gyr),
one expects a significant amount of new material (gas and dark matter)
to be accreted in a cosmological framework, but this accretion has
been neglected in most binary merger studies. This accretion can
greatly affect the evolution of the galaxies
\citep{moster2011a}. Furthermore, many galaxies experience multiple
mergers over their lifetime, and indeed, these mergers frequently
occur in rapid enough succession that the galaxy does not have time to
relax in between (see section \ref{sec:mwmergers}). Therefore this
process may not be accurately simulated via a sequence of binary
mergers. 

\subsection{Semi-analytic models}

The semi-analytic approach was proposed by \citet{white1991}, based on
ideas presented in \citet{white1978}. This method, which could only
predict average quantities, was later extended to model the properties
of individual systems, based on merger histories as predicted by the
CDM scenario
\citep{kauffmann1993,cole1994,kauffmann1994,baugh1996,somerville1999a,bower2006,
  croton2006,somerville2008a}. In a semi-analytic model (SAM) one
starts by specifying a cosmological model and then traces the merger
histories for a series of dark matter haloes using either an $N$-body
simulation or the analytic extended Press-Schechter (EPS)
formalism. The evolution of the baryonic component in the haloes is
followed by using simple, yet physical analytic recipes for gas
cooling and accretion, star formation and feedback. Cooling converts
hot gas into cold gas, star formation converts cold gas into stars,
and feedback converts cold gas into hot gas and in some cases ejects
it from the halo.  When two haloes merge, it is assumed that the
galaxies merge on a time scale set by dynamical friction, possibly
altering the properties of both systems. A sample of model galaxies
that represents the galaxy population in the present-day Universe can
then be built by modelling a large number of haloes that sample the
halo mass function. The resulting star formation and chemical
evolution histories can be convolved with stellar population synthesis
models to predict observables such as luminosity and colour. For an
excellent recent review on SAMs we refer to \citet{baugh2006}.

Thus, in a SAM the complicated astrophysical processes that are
responsible for the formation and evolution of galaxies are modelled
as a set of recipes which carry a number of free parameters (it should
be kept in mind, however, that numerical hydrodynamic simulations
carry the same free parameters to characterize the sub-grid
physics). As the current understanding of many of these processes is
limited, the model parameters cannot be derived from first
principles. Instead, the model is normalised with a set of
observational constraints, and with the help of more detailed
hydrodynamical simulations. The advantage of SAMs is their flexibility
and their computational efficiency. It is possible to run many
realisations in a short amount of time, and thereby test the effects
of the various assumptions and model parameters. It is also possible
to create `mock catalogs' of large samples of galaxies (millions),
which can be compared with large observational samples. Moreover, one
can include more physical processes, such as AGN feedback, albeit in a
schematic way. Currently, SAMs are much more successful than numerical
cosmological simulations at reproducing the global properties of
galaxies. Furthermore, SAMs reproduce quite well the results of
hydrodynamic ``zoom'' simulations when similar physics is included
\citep{hirschmann2012}. 

The main drawback of SAMs is that they yield only approximate and
schematic information about the internal structure and morphology of
galaxies. For example, most SAMs provide an estimate of the ratio of
the spheroid mass to the disc mass, based on the merger history of the
galaxy and the assumption that near-equal mass mergers destroy discs
and build spheroids \citep[e.g.][]{fontanot2011}.
Many SAMs also include a simple recipe for
computing the radial size and rotation velocity of the disc component
based on angular momentum considerations
\citep[e.g.][]{somerville2008a}, and recent work provides an approach
for predicting the size and velocity dispersion of spheroids formed in
mergers \citep{covington2011}. However, in the existing SAMs, these
recipes are largely based on binary merger simulations, which we have
argued may not accurately capture the associated physics because of the
neglect of the cosmological accretion. In summary, we may consider the
SAM to be an efficient tool for predicting the global properties of
large samples of galaxies, but of limited use for studying galaxy
internal properties.

\subsection{Combining simulations and semi-analytic models}

In this paper, we present a new approach that leverages the
complementary strengths of semi-analytic models and hydrodynamic
merger simulations to efficiently simulate the evolution of galaxies
at high resolution and within a cosmological context. We start with a
dark matter halo merger tree extracted from a dissipationless N-body
simulation. We run the SAM within this halo merger tree to construct a
\emph{galaxy} merger tree, which specifies when each galaxy enters a
larger halo and becomes a satellite, and the global properties of this
galaxy (e.g. stellar mass of a bulge and disc component, cold gas
mass, radial size) when it enters the halo. We use these SAM
predictions to specify the initial conditions for a sequence of
mergers, which we simulate with a full numerical hydrodynamic code.
Specifically, we evolve the galaxy in the main branch with our hydrodynamical code,
and include satellite galaxies in the simulation at the time when
they cross the virial radius of the larger halo (with this time specified by the
Nbody similulation) using the satellite properties predicted by the SAM. 
In addition, we include the growth of the main
halo due to cosmological accretion, as specified by the merger tree. 

This approach is particularly well suited for studies that focus on the
evolution of the internal structure of galaxies, including different
components such as a thin and thick disc, spheroid, and stellar halo.
While many applications of our model focus on the evolution of the
central galaxy, we can also address the evolution of the satellite
population. Finally, it is our goal to develop generalised
semi-analytic prescriptions based on the results of our
simulations. In this way, the existing semi-analytic recipes can also
be improved.


\section{Methods}
\label{sec:methods}

In this section we briefly describe the tools that are used in this
paper. These are the simulation code {\sc GADGET-2} that was employed
both for the cosmological $N$-body simulation from which our merger
trees were extracted and the hydrodynamic merger simulations, the code
to create initial galaxy models, and the semi-analytic model used to
populate N-body merger trees with galaxies. Throughout this paper we
adopt cosmological parameters chosen to be consistent with results
from {\sc WMAP}-3 \citep{spergel2007} for a flat $\Lambda$CDM
cosmological model: $\Omega_m=0.26$, $\Omega_{\Lambda}=0.74$,
$h=H_0/(100$~km~s$^{-1}$~Mpc$^{-1})=0.72$, $\sigma_8=0.77$ and
$n=0.95$. These parameters differ only slightly from the currently
favored WMAP7 parameters \citep{komatsu2011}. We adopt a
\citet{kroupa2001} IMF and compute all stellar masses accordingly.

\subsection{N-body and Smoothed Particle Hydrodynamics code}
\label{sec:ncode}

For all simulations in this work we employ the parallel TreeSPH-code
{\sc GADGET-2} \citep{springel2005a}.  The code uses Smoothed Particle
Hydrodynamics \citep[SPH;][] {lucy1977,gingold1977,monaghan1992} to
evolve the gas using an entropy conserving scheme
\citep{springel2002}. Radiative cooling is implemented for a
primordial mixture of hydrogen and helium following \citet{katz1996},
and a spatially uniform time-independent local UV background in the
optically thin limit \citep{haardt1996} is included. All simulations
have been performed with a high force accuracy of $\alpha_{\rm
  force}=0.005$ and a time integration accuracy of $\eta_{\rm
  acc}=0.02$ \citep[for further details see][]{springel2005a}.

Star formation and the associated heating by supernovae (SN) is
modelled following the sub-resolution multiphase ISM model described
in \citet{springel2003}. The ISM in the model is treated as a
two-phase medium with cold clouds embedded in a hot component at
pressure equilibrium.  Cold clouds form stars in dense
($\rho>\rho_{th}$) regions on a timescale chosen to match observations
\citep{kennicutt1998}. The threshold density $\rho_{th}$ is determined
self-consistently by demanding that the equation of state (EOS) is
continuous at the onset of star formation. SN-driven galactic winds
are included as proposed by \citet{springel2003}. In this model the
mass-loss rate carried by the wind is proportional to the star
formation rate (SFR) $\dot M_w= \eta \dot M_*$, where the
mass-loading-factor $\eta$ quantifies the wind
efficiency. Furthermore, the wind is assumed to carry a fixed fraction
of the supernova energy, such that there is a constant initial wind
speed $v_w$ (energy-driven wind). We do not include feedback from
accreting black holes (AGN feedback) in our simulations.

We compute the parameters for the multiphase feedback model following
the procedure outlined in \citet{springel2003} in order to match the
Kennicutt Law. For a Kroupa IMF the mass fraction of massive stars is
$\beta=0.16$ resulting in a cloud evaporation parameter of $A_0=1250$
and a SN ``temperature'' of $T_{\rm
  SN}=1.25\times10^8{\rm~K}$. Finally, the star formation timescale is
set to $t_*^0=3.5{\rm~Gyr}$. For the galactic winds we adopt a
mass-loading factor of $\eta = 1$ and a wind speed of $v_w \sim 500
\kms$.

\subsection{Merger Trees}
\label{sec:trees}

For this study we use dark matter merger trees drawn from an $N$-body
simulation run with the {\sc GADGET-2} code.  The initial conditions
for the {\sc WMAP}-3 cosmology were generated using the {\sc GRAFIC}
software package \citep{bertschinger2001}.  The simulation was done in
a periodic box with a side length of $100$ Mpc, and contains $512^3$
particles with a particle mass of $2.8\times 10^8\Msun$ and a comoving
force softening of $3.5$ kpc. Starting at a redshift of $z=43$, 94
snapshots were stored until $z=0$, equally spaced in expansion factor
($\Delta a=0.01$).

Dark matter haloes were identified in the simulation snapshots using a
Friends of Friends (FOF) halo finder with a linking parameter of
$b=0.2$. Substructures inside the FOF groups are then identified using
the {\sc SUBFIND} code. For the most massive subgroup in a FOF group
the virial radius and mass are determined with a spherical overdensity
criterion using the fitting function by \citet{bryan1998}. The minimum
particle number for haloes is set to 20, resulting in a minimum halo
mass of $5.6\times10^9\Msun$. The number of parent haloes found at
$z=0$ with this method is $\sim135~000$ and parent halo masses range
between $10^{10}\Msun$ and $10^{15}\Msun$. The merger trees are
constructed for all parent haloes at $z=0$ by connecting haloes
between the 94 catalogues. The branches of the trees were determined
by linking every halo to its most massive progenitor at previous
snapshots. In total, we have 41~000 merger trees.

\subsection{Galaxy models}
\label{sec:models}

To construct the galaxy models used in our simulations we apply the
method described in \citet{springel2005b} with the extension by
\citet{moster2011a}. Each system is composed of a cold gaseous disc, a
stellar disc and a stellar bulge with masses \Mcg, \Mdisc and \Mb
embedded in a halo that consists of hot gas and dark matter with
masses \Mhg and \Mdm.

The gaseous and stellar discs are rotationally supported and have
exponential surface density profiles.  The scale length of the gaseous
disc \rg is related to that of the stellar disc \rd by $\rg = \chi
\rd$. The vertical structure of the stellar disc is described by a
radially independent sech$^2$ profile with a scale height $z_0$, and
the vertical velocity dispersion is set equal to the radial velocity
dispersion. The gas temperature is fixed by the EOS, rather than the
velocity dispersion. The vertical structure of the gaseous disc is
computed self-consistently as a function of the surface density by
requiring a balance of the galactic potential and the pressure given
by the EOS. The spherical stellar bulge is non-rotating and is
constructed using the \citet{hernquist1990} profile with a scale
length \rb. The dark matter halo has a \citet{hernquist1990} profile
with a scale length \rs, a concentration parameter $c=\rvir/\rs$ and a
dimensionless spin parameter $\lambda$.

The hot gas is modelled as a slowly rotating halo with a spherical
density profile following \citet{moster2011a}.  We employ the
observationally motivated $\beta$-profile
\citep{cavaliere1976,jones1984,eke1998}:
\begin{equation}
\rho_{\rm hg}(r) = \rho_0 \left[1+\left(\frac{r}{\rc}\right)^2\right]^{-\frac{3}{2}\beta}\;,
\end{equation}
which has three free parameters: the central density $\rho_0$, the core radius \rc~and the outer slope parameter
$\beta$. We adopt $\beta = 2/3$ \citep{jones1984}, $\rc=0.22\rs$ \citep{makino1998} and fix $\rho_0$ such that the hot gas mass within
the virial radius is $M_{\rm hg}$.

The temperature profile is fixed by assuming an isotropic model and
hydrostatic equilibrium inside the galactic potential such that it is
supported by pressure. In addition, the hot gaseous halo is rotating
around the spin axis of the disc with a specific angular momentum
$j_{\rm hg}$, which is a multiple of the specific angular momentum of
the dark matter halo $j_{\rm dm}$ such that $ j_{\rm hg} = \alpha
j_{\rm dm}$. The angular momentum distribution is assumed to scale
with the product of the cylindrical distance from the spin axis $R$
and the circular velocity at this distance: $j(R) \propto R \; v_{\rm
  circ}(R)$.

High resolution cosmological simulations that succeed in reproducing
disc sizes find that at low redshift ($z\lta2$), $\alpha$ is greater
than unity. This is because feedback processes preferentially remove
low angular momentum material from the halo \citep{governato2010}.
Similarly, \citet{moster2011a} constrain $\alpha$ by using isolated
simulations of a MW-like galaxy and demanding that the evolution of
the average stellar mass and scalelength found observationally be
reproduced. The model that agrees best with the observational
constraints has a spin factor of $\alpha=4$. We use this value
throughout this work.

\subsection{Semi-analytic model}

In this paper we use the SAM constructed by \citet{somerville1999a} in
its current implementation \citep{somerville2008a,somerville2012}. In
this model each dark matter halo is assigned two properties: the spin
parameter $\lambda$ and the concentration parameter $c$. Each
top-level halo is assigned a value of $\lambda$ by selecting values
randomly from a lognormal distribution with mean $\bar \lambda=0.05$
and width $\sigma_{\lambda}=0.5$ \citep{antonuccio2010}.  The initial
density profile of each halo is described by the NFW form. For
$N$-body merger trees the concentration parameter and the halo
position are extracted from the simulation.  After accretion the
orbital evolution of each subhalo is computed using the dynamical
friction formula given by \citet{boylankolchin2008}, including the
effects of tidal stripping and destruction.

The SAM adopts a simple but fairly standard spherical cooling model,
which makes use of the multi-metallicity cooling function of
\citet{sutherland1993}.
The hot gas density profile is assumed to be that of a singular
isothermal sphere. In order to model the effects of a photo-ionising
background, the SAM follows \citet{gnedin2000} and
\citet{kravtsov2004} and defines a \lq filtering mass\rq~$M_{\rm F}$
which is a function of redshift and depends on the re-ionisation
history of the Universe.  The cold gas profile is assumed to be an
exponential disc with a scalelength proportional to the scalelength of
the stellar disc. Only gas lying above a critical surface density
threshold is available for star formation. In the \lq
quiescent\rq~phase star formation is based on the empirical
Schmidt-Kennicutt law with the appropriate normalisation for a Kroupa
IMF.  Massive stars and supernovae impart thermal and kinetic energy
to the cold interstellar medium, and cold gas may be ``reheated'' from
the disc and either deposited in the hot halo or ejected from the halo
altogether.

When a subhalo loses its orbital energy due to dynamical friction the
satellite merges with the central galaxy, driving a starburst. The
efficiency of star formation in a merger-triggered burst is
parametrized as a function of the mass ratio of the merging pairs and
the gas fraction of the progenitors and calculated with the model
proposed by \citet{hopkins2009a}, which is based on binary merger
simulations. It is assumed that during every merger with a mass ratio
above $\mu=0.1$, a fraction of the disc stars is transferred to the
spheroidal or bulge component, again following the prescription
outlined in \citet{hopkins2009a}.

The SAM contains recipes for black hole growth and AGN feedback due to
both ``bright mode'' AGN-driven winds and ``radio mode'' heating by
radio jets \citep{somerville2008a}. However, at the halo mass scale
considered here, AGN feedback has little impact on the predictions.


\section{Multiple mergers in a $\Lambda$CDM universe}
\label{sec:mwmergers}

In order to help motivate this work we first address whether most
mergers can be treated as ``isolated'' or whether multiple mergers are
expected to be cosmologically significant. In an \lq
isolated\rq~merger (also called \lq binary\rq~merger), two galaxies
merge and become dynamically relaxed before the remnant merges with
another galaxy. In \lq multiple\rq~mergers, two galaxies have not had
enough time to merge and become relaxed when another galaxy already
enters their common halo. If isolated mergers are the standard event,
then it is possible to study single merger events with specified
parameters (e.g. in simulations) and then \lq stack\rq~these merger
events for a given merger history, in order to determine the effects
of the mergers on the galaxy. This can be seen as a linear
process. However, if multiple mergers are more common, then it may not
be possible to string the results from binary merger simulations
together. One then has to study the events with three or more galaxies
involved which will have a different impact on the galaxy properties
than two or more binary mergers. The merger process then becomes
non-linear and the (already large) parameter space for the merger
parameters multiplies.

\begin{figure*}
\psfig{figure=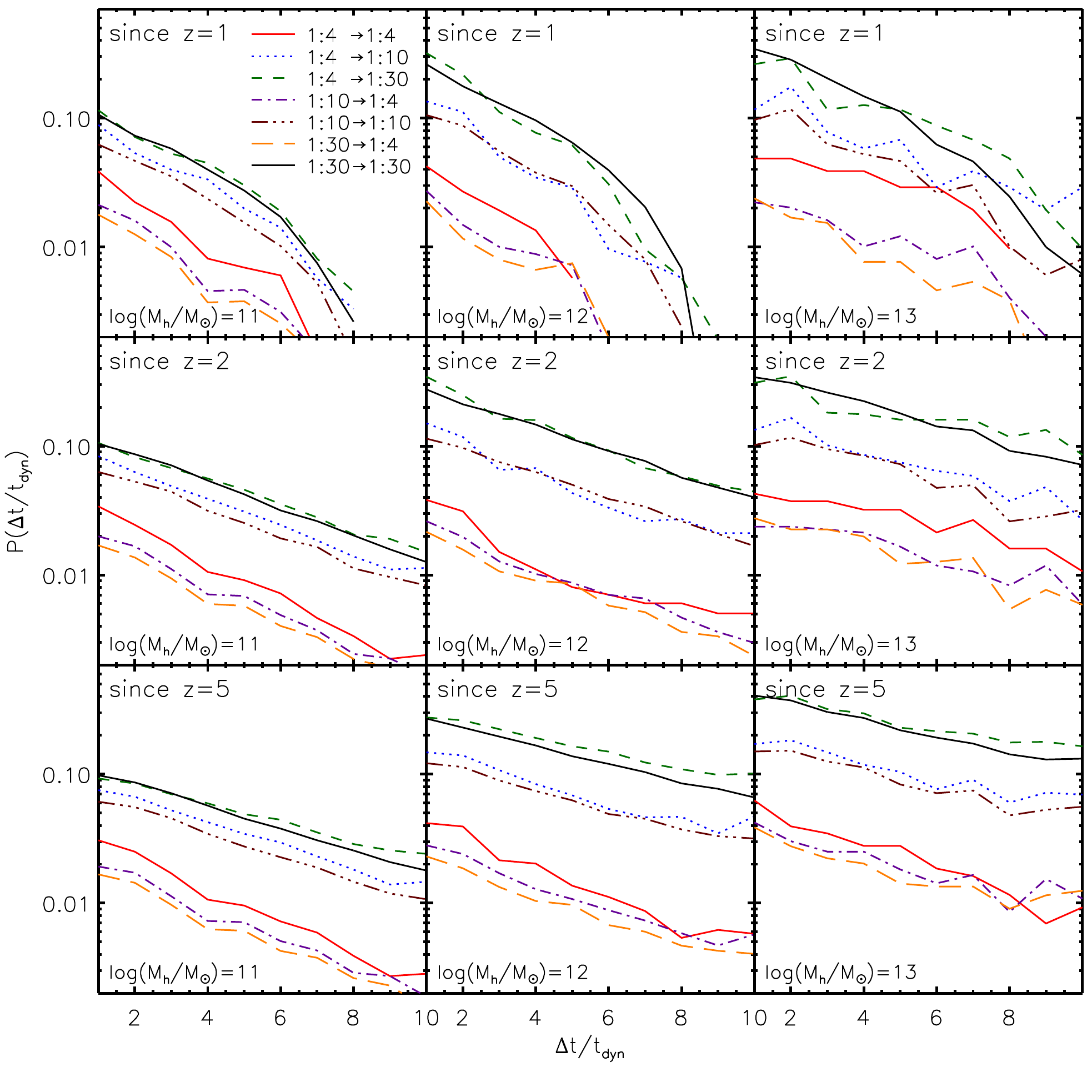,width=0.95\textwidth}
\caption{Probability for multiple mergers. The lines give the
  probability for a sequence of mergers with the indicated mass ratios
  as a function of the time difference between the mergers in units of
  the dynamical time of the parent halo.}
\label{fig:ptdyn}
\end{figure*}

In order to study whether binary or multiple mergers are more common,
we make use of the cosmological merger trees described in
Section~\ref{sec:trees}.  We divide the merger trees according to the
mass of their parent halo at $z=0$ with a bin size of $\Delta
\log(M/\Msun)=0.5$. We then identify all merger trees of these systems
with two or more halo mergers and for every pair of mergers we record
the time difference between the halo mergers. If e.g. the halo of the
satellite galaxy A merges with the halo of the central galaxy C at a
cosmic time $t_1=8.0\Gyr$ and later the halo of the satellite galaxy B
merges with the halo of the central galaxy at $t_2=10.0\Gyr$, we store
the time difference $\Delta t=2.0\Gyr$. We divide this time by the
dynamical time of the halo $t_{\rm dyn}=r_{\rm vir}/V_{\rm vir}$ at
the time of the first halo merger $t_1$ and distribute the resulting
number in bins of width $\Delta t/t_{\rm dyn}=1$. Finally, we count
the number of mergers pairs with a given mass ratio sequence and since
a given redshift for each time bin. The sequence 1:4 $\rightarrow$
1:10 for example means that the first merger has a mass ratio above
1:4 and the second merger has a mass ratio above 1:10. In order to get
the probability for a merger sequence we normalise this number by
dividing it by the number of mergers with the mass ratio of the first
merger.

We show the results of this analysis in Figure \ref{fig:ptdyn}. The
probability for two mergers happening within a few dynamical times of
the halo is always higher than the probability for a larger time
difference. This indicates that usually the second satellite galaxy
enters the parent halo before the first satellite had time to merge
with the central galaxy, which happens only after several dynamical
times, depending on the mass ratio of the merger. For the mergers that
happen since $z=1$ (upper row in the figure) his trend is most
obvious. Here, the second satellite enters the halo within three
dynamical times after the first satellite entered for more than 50 per
cent of all merger pairs independent of mass ratio. If we start at a
higher redshift, the merger pairs are on average a little more
separated. Still for more the 50 per cent of all merger pairs, the
second satellite enters the halo within five dynamical times after the
first one since $z=5$.

In summary, we find that multiple mergers are more common than
sequences of isolated binary mergers. Therefore it is not optimal to
focus on binary mergers and afterwards try to string their effects
together.  Thus merger simulations which consider several satellites
entering the halo and orbiting together within the halo have to be
performed and analysed.  If parameters for the galaxies and the orbits
are chosen from a multidimensional grid, the parameter space is then
too large to cover. This means that one has to reduce this huge
parameter space by selecting only those parameters that are common for
the chosen cosmology. $N$-body merger trees populated with galaxies
using a SAM offer this option and are therefore very practical to use
as initial conditions for the merger simulations. In this way, mergers
that are common in the Universe are automatically selected. Thus, the
important regions of the parameter space are naturally covered.

\section{Simulations of galaxy merger trees} 
\label{sec:smt}

Galaxy merger trees obtained from $N$-body merger trees in combination
with a SAM provide natural initial conditions for merger simulations
with high resolution. However, if we compare the galaxy models
predicted by the SAM to the model galaxies used in merger simulations,
we immediately recognise an important difference: the standard
approach in binary merger simulations neglects the accretion of
material coming from outside the initial virial radius. Standard codes
used to create model galaxies only employ a dark matter halo of fixed
mass.  This means that all material that falls into the halo which is
not included in the merger tree (i.e. does not come through
satellites) is not accounted for.

On the other hand, cosmological N-body simulations show a substantial
increase of the mass of galactic dark matter haloes through ``smooth
accretion'', that then must be taken into account in a cosmologically
based study of galaxy evolution. We now describe our method for
including this smooth accretion in our simulations.

\subsection{Modelling the growth of the dark halo}
\label{sec:smooth}

In order to construct a merger tree one must define a lower mass limit
(mass resolution) below which the tree is truncated.  This means that
all haloes that are smaller than this mass limit and fall into a
larger halo are not considered as merging haloes, but as so-called \lq
smooth accretion\rq. Thus, even if there are no mergers in a given
time-step, the parent halo gains mass. In (non-cosmological)
simulations of galaxies or mergers this smooth accretion is usually
neglected.

We model this accretion by placing the additional dark matter
particles into small spherical systems which will be denoted as \lq
DM-spheres\rq~in the following.  We place all DM-spheres at a specific
distance from the halo centre such that the accretion rate of the
DM-spheres matches the smooth accretion rate of the merger tree. The
DM-spheres are uniformly distributed around the halo. In order to
prevent the DM-spheres from falling to the very centre of the halo and
thereby disturbing the disc, we give the DM-spheres an initial
velocity in a random direction orthogonal to the radius vector. This
of course, reduces the initial distance to the centre compared to
freely falling particles. As we require a density profile that leads
to a quick dissolving of the DM-sphere once it enters the halo (since
the DM-spheres are not meant to represent substructure but smooth
accretion) we employ a three dimensional Gaussian density
profile. Note, that we do not add gas particles to the DM-spheres.

\subsubsection{Creating a spherical particle distribution}

We model each DM-sphere with a three dimensional Gaussian density
profile
\begin{equation}
\rho(r) = \frac{M_{\rm DMS}}{\left(\sqrt{2\pi}r_{\rm DMS}\right)^3}\exp\left(-\frac{r^2}{2r_{\rm DMS}^2}\right)\,,
\end{equation}
where $M_{\rm DMS}$ is the total mass of the DM-sphere and $r_{\rm
  DMS}$ is a scalelength. The mass $M(r)$ enclosed within a radius $r$
is then found by integrating the density profile and the gravitational
potential is found by integrating Poisson's equation. Using $N$
particles of mass $m_i$ for each DM-sphere, a Gaussian density profile
can by simply achieved by drawing a random number from a Gaussian
distribution with width $r_{\rm DMS}/\sqrt3$ for each cartesian
coordinate. Deriving the velocity for each particle is more
complicated and has to be done using the distribution function as a
function of the relative energy $f(\mathcal{E})$ which can be obtained
with the so-called \lq Eddington inversion\rq~\citep[see
  e.g.][]{binney1987}.

In order to solve for $f(\mathcal{E})$, we create a logarithmically
spaced array in radius with $10^5$ bins. We fix the minimum and
maximum of the bins at $10^{-3}$ and $10^2$ times the scalelength. On
this finely spaced grid, we define the values for $\rho(r)$, $M(r)$ and
$\Psi(r)$ and obtain the derivatives by finite differencing. The
distribution function is then obtained by numerically integrating the
Eddington equation. We find that it is well approximated by the
fitting function
\begin{eqnarray}
f(\mathcal{E}) &=& \frac{2.2\times10^{-3}\Msun}{(\kpc\kms)^3} \; \sqrt{\frac{\Msun\kpc^3}{M_{\rm DMS}r_{\rm DMS}^3}}\nonumber\\
&& \times Q~\left[\left(\frac{Q}{0.3}\right)^{-3.871}+\left(\frac{Q}{0.3}\right)^{0.166}\right]^{-5.4}\,,
\end{eqnarray}
where $Q=\mathcal{E}/\Psi_0$ and $\Psi_0=-\Phi_0=\sqrt{2/\pi}GM_{\rm
  DMS}/r_{\rm DMS}$ is the maximum relative potential. With the
distribution function, we can now find the velocity for each particle
with a rejection sampling technique. The direction of the velocity is
randomly chosen from the unit sphere.

\begin{figure}
\psfig{figure=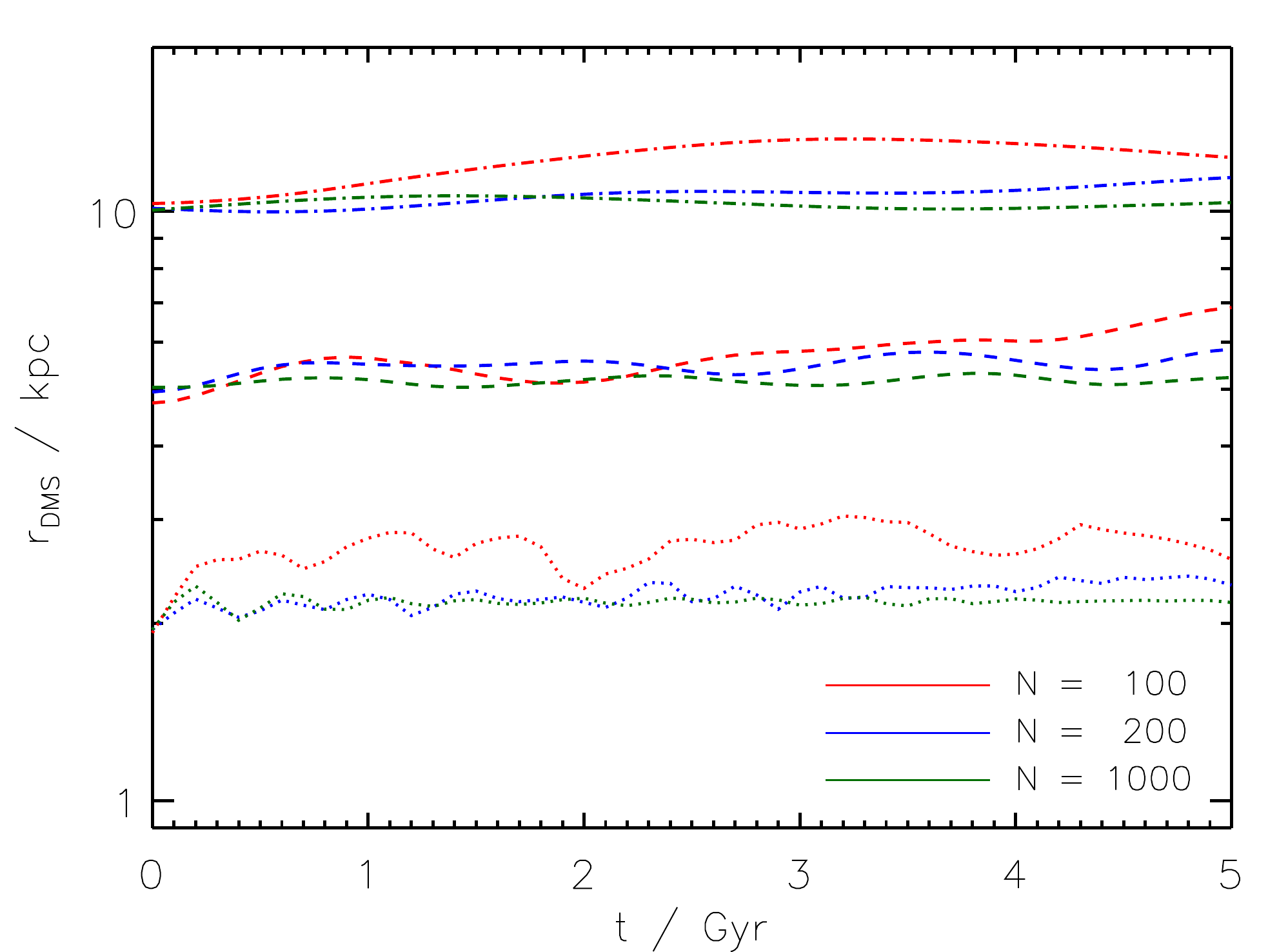,width=0.45\textwidth}
\caption{Evolution of the scalelength of a Gaussian DM-sphere in
  simulations with 100, 200 and 1000 particles. Initial scalelengths
  are 2, 5 and 10 $\kpc$ represented by the dotted, dashed and
  dash-dotted lines, respectively.}
\label{fig:dmsevo}
\end{figure}

We test this model by creating $N$-body realisations of a DM-sphere of
mass $M_{\rm DMS}=10^9\Msun$. This mass is below the resolution limit
of the cosmological simulation used in section \ref{sec:mwmergers} and
is thus accounted for as smooth accretion.  We model the DM-sphere
with $N=100, 200$ and $1000$ particles and a scalelength of $r_{\rm
  DMS}=2, 5$ and $10\kpc$. We let each realisation evolve in isolation
for $5\Gyr$ and measure the scalelength at every time-step. The
results of this analysis are shown in Figure \ref{fig:dmsevo}. All
models are stable over the $5\Gyr$ of evolution, independent of the
initial scalelength. For $N=100$ the scalelength increases by about 50
per cent until the end of the simulation, while for larger particle
numbers the scalelength deviates only slightly from the initial
value. For all particle numbers, models with a larger initial
scalelength are more stable.

\subsubsection{Placing the dark matter systems around the halo}

With the model presented in the last section we can create DM-spheres
that are stable in isolation, but due to their shallow
potential they dissolve quickly when orbiting a dark matter halo. The
next step in modeling the smooth accretion of a dark matter halo is to
place dark matter at the right position, such that the accretion
history of the halo is reproduced. Thus, the main difficulty here is
where to position the DM-spheres and with which mass.

\begin{figure*}
\psfig{figure=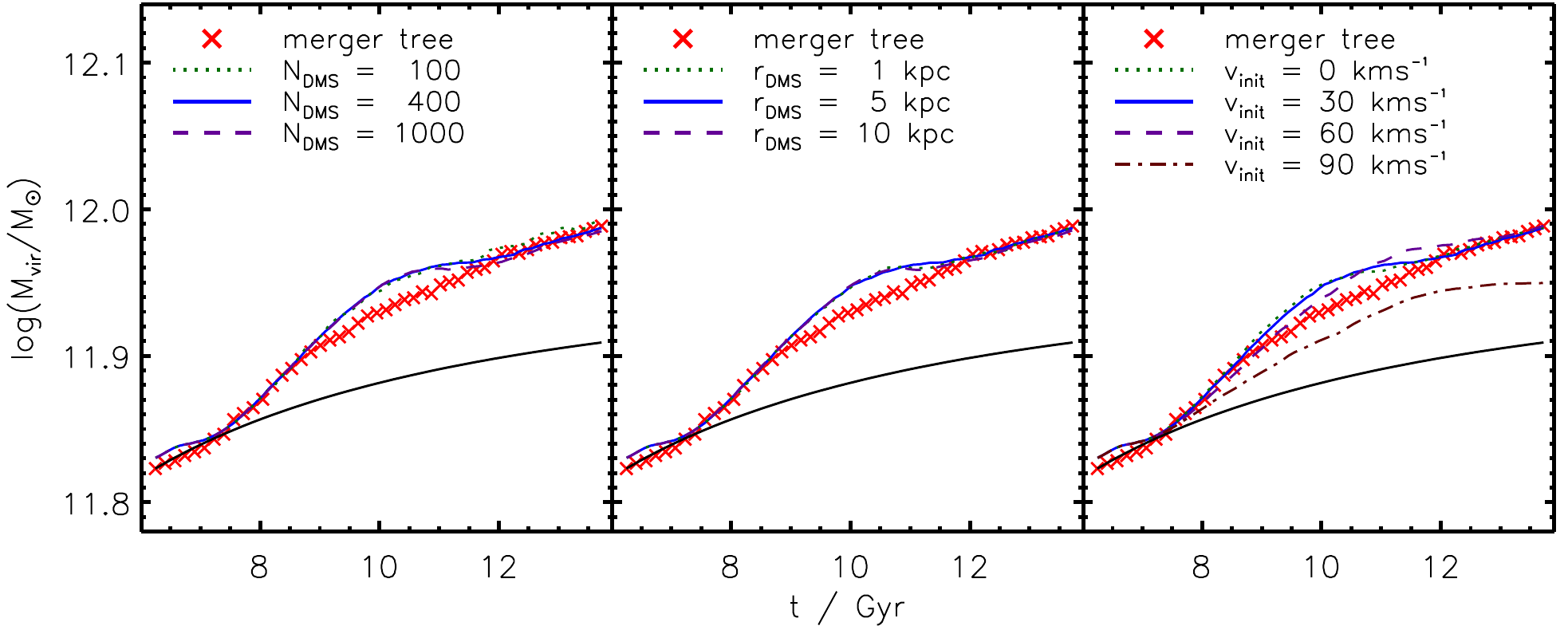,width=0.95\textwidth}
\caption{Growth of the halo mass due to smooth accretion from
  DM-spheres. The virial mass measured in the isolated simulations
  (lines) is compared to that found in the merger tree (symbols). In
  the left, middle and right panels different numbers of DM-spheres,
  scalelengths and initial velocities have been used,
  respectively. Unless otherwise noted, the fiducial values $N_{\rm
    DMS}=400$, $r_{\rm DMS}=5\kpc$ and $v_{\rm init}=30\kms$ have been
  used. The black line shows the mass evolution with no smooth
  accretion included.}
\label{fig:dms}
\end{figure*}

As we want to simulate merger trees, we first extract the mass
accretion history of the \lq main branch\rq~from the $N$-body
simulation, i.e. the mass of the main halo and its most massive
progenitors as a function of cosmic time. The next step is to choose a
starting redshift $z_{\rm start}$ which corresponds to a starting time
$t_{\rm start}$, at which we start the simulation. If we assume that
the profile of the parent halo does not change, which is true for our
initial conditions, the virial mass of the parent halo increases as
the background density decreases towards lower redshift (for a lower
$\rho_{\rm crit}$ the radius that contains a certain overdensity
becomes larger). We thus subtract the increasing virial radius at
every time-step such that we are left with the accreted
mass. Furthermore, we subtract the mass that has been accreted through
mergers of resolved smaller haloes, as those will be explicitly
simulated as merging haloes. We are thus left with the mass $M_{\rm
  smooth}(t)$ that has been added to the parent halo due to smooth
accretion.

In the next step we choose the number of DM-spheres $N_{\rm DMS}$,
which we set equal to the number of time-steps we will use for the
model. We then interpolate $M_{\rm smooth}(t)$ for every time-step
$t_i$ using a Bezier curve which traces the mass accretion history and
is equal to the mass found in the merger tree at $t_{\rm start}$ and
$t_{N_{\rm DMS}}$. We then compute the mass that is accreted within
every time-step $M_{\rm DMS}(t_i)$, and use this mass for the
DM-sphere that corresponds to this time-step. If this mass is smaller
than zero (i.e. the parent halo loses mass), we set the mass of the
corresponding DM-sphere to zero (i.e. the DM-sphere is
omitted). However, we store this lost mass and set the mass of the
DM-spheres that correspond to the following time-steps to zero, until
the lost mass is balanced by accreted mass. 
Thus the overall mass growth via smooth accretion over the whole
accretion history is reproduced. Finally, we set
the particle mass for each DM-sphere particle equal to the mass of the
dark matter halo particles $m_{\rm dm}$ and compute the number of
particles of each DM-sphere: $N_i = M_{\rm DMS}(t_i)/m_{\rm dm}$.

Having determined the mass of each DM-sphere, we now need to specify
where to place them. For this we make use of the free-fall time for a
point mass at a radius $r$: $t_{\rm ff} = \pi [r^3/8GM(r)]^{1/2}$,
where $M(r)$ is the total mass enclosed within $r$. The time it takes
a DM-sphere to enter the virial radius of the halo $t_{\rm enter}=
t_i-t_{\rm start}$ is shorter than the free-fall time and only
slightly larger than the free-fall time at radius $r$ minus the
free-fall time at the virial radius $r_{\rm vir}$, as the DM-sphere
already has a velocity towards the centre when crossing the virial
radius. We can thus assume that $t_{\rm enter}$ has a value of $t_{\rm
  enter}\gta t_{\rm ff}(r_i)-t_{\rm ff}(r_{\rm vir})$, where $r_i$ is
the initial distance of the DM-sphere from the halo centre and $r_{\rm
  vir}$ is the virial radius of the halo at $t_i$.  The initial
distance of each DM-sphere is then given by
\begin{equation}
r_i \approx \left(\frac{\sqrt{8GM(r)}}{\pi}~t_{\rm enter} + r_{\rm vir}^{3/2}\right)^{2/3}\,,
\end{equation}
and we place the centre of each DM-sphere randomly on a sphere with
radius $r_i$. In order to prevent the DM-spheres from passing through
the centre of the halo, we assign an initial velocity to every
DM-sphere in a random direction orthogonal to the radius vector. This
leads to a larger time taken for a DM-sphere to enter the virial radius which
partly compensates for smaller estimate of $t_{\rm enter}$.

We are thus left with three free parameters: the number of DM-spheres
$N_{\rm DMS}$, their scalelength $r_{\rm DMS}$, and their initial
velocity $v_{\rm init}$. In order to fix these values we select a
merger tree from the simulation box, presented in section
\ref{sec:mwmergers}, that has no mergers above a mass ratio
$\mu=0.02$. We can thus assume that all mass accreted by the parent
halo is from smooth accretion. The index number of this merger tree is
\textit{1811} and the virial mass at $z=0$ is $10^{12}\Msun$. We start
the simulation at $z_{\rm start}=1$, where the parent halo has a
virial mass of $6.5\times10^{11}\Msun$, a virial velocity of $V_{\rm
  vir}=150\kms$, We construct an $N$-body realisation of this dark
matter halo with the initial conditions generator as described in
section \ref{sec:models}, but omitting all baryonic components. We
employ $N_{\rm halo}=200~000$ particles, and add dark matter particles
in DM-spheres to this model with the method outlined above. For this
we choose the fiducial parameters as $N_{\rm DMS}=400$, $r_{\rm
  DMS}=5\kpc$ and $v_{\rm init}=30\kms$, and vary each parameter
around these values, while keeping the other two parameters fixed. For
about 400 DM-spheres the number of particles per DM-sphere is
$\approx200$ which we have shown to be sufficient for stable
DM-spheres in isolation. We evolve each system until $z=0$ with a
softening length of $\epsilon=0.7\kpc$.

\begin{figure}
\includegraphics[width=0.45\textwidth]{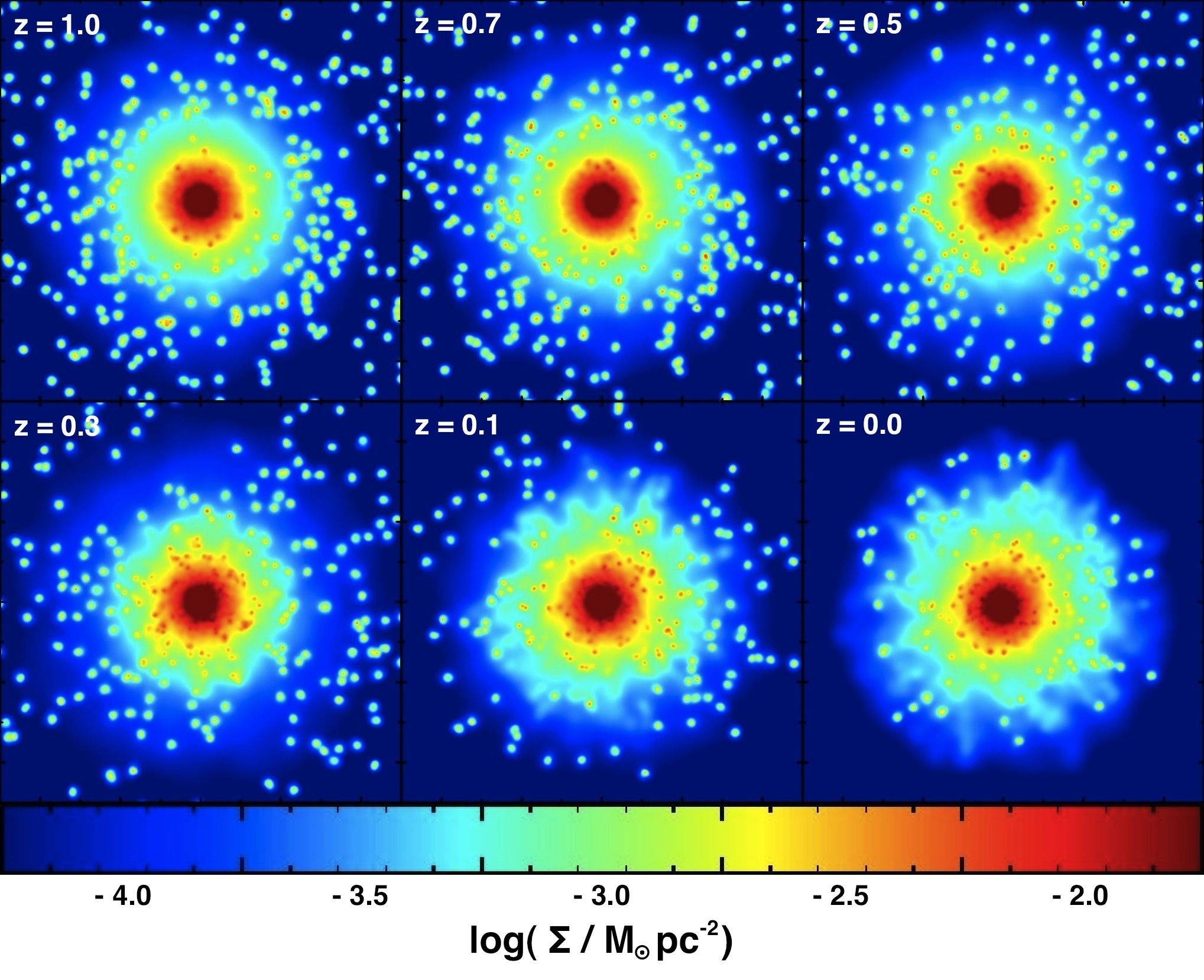}
\caption{Projected dark matter surface density for the smooth
  accretion model using the fiducial parameters. Blue colour
  represents regions of low density and red colour depicts high
  densities. Each panel measures 1 Mpc on a side and the redshift is
  displayed in the upper left corner of each panel. The virial radius
  of the halo is given by the envelope of the green region.}
\label{fig:dms1mpc}
\end{figure}

For each simulation we measure the virial mass as a function of cosmic
time using a spherical overdensity criterion and the fitting function
by \citet{bryan1998}. The results are presented in Figure
\ref{fig:dms}, where the virial mass measured in the simulations is
compared to that found in the merger tree. For our fiducial
parameters, we are able to reproduce the accretion history of the halo
very well. In the left panel the number of DM-spheres is varied. We
see that neither a lower nor a higher value than the fiducial one
results in a difference for the mass that can be accreted. In the
middle panel we vary the scalelength of the Gaussian profile. Also for
this parameter we see that the result is independent of its value. In
the right panel the initial velocity is varied. Values of $v_{\rm
  init}=0$ and $30\kms$ lead to a very similar mass accretion
history. For $v_{\rm init}=60\kms$ the DM-spheres enter the virial
radius slightly later, as the orbital energy is higher and the
DM-spheres need to lose some angular momentum first. For higher values
($v_{\rm init}\gta90\kms$) the virial mass of the halo is too low,
compared to the merger tree. The reason is that for some DM-spheres
the initial velocity is high enough, such that they orbit within the
virial radius of the parent halo just for a short time, or even not at
all. Instead their pericentric distance has the same dimension as the
virial radius. Thus they do not fall towards the centre, but orbit the
halo on a large distance where the density is not high enough such
that they can lose their angular momentum due to dynamical
friction. This means that the initial velocities of the DM-spheres
should not be larger than half of the virial velocity of the
halo. Altogether we find that our model is able to reproduce the
smooth accretion history of a dark matter halo very well and does not
depend on the exact values of the model parameters.

\begin{figure}
\includegraphics[width=0.45\textwidth]{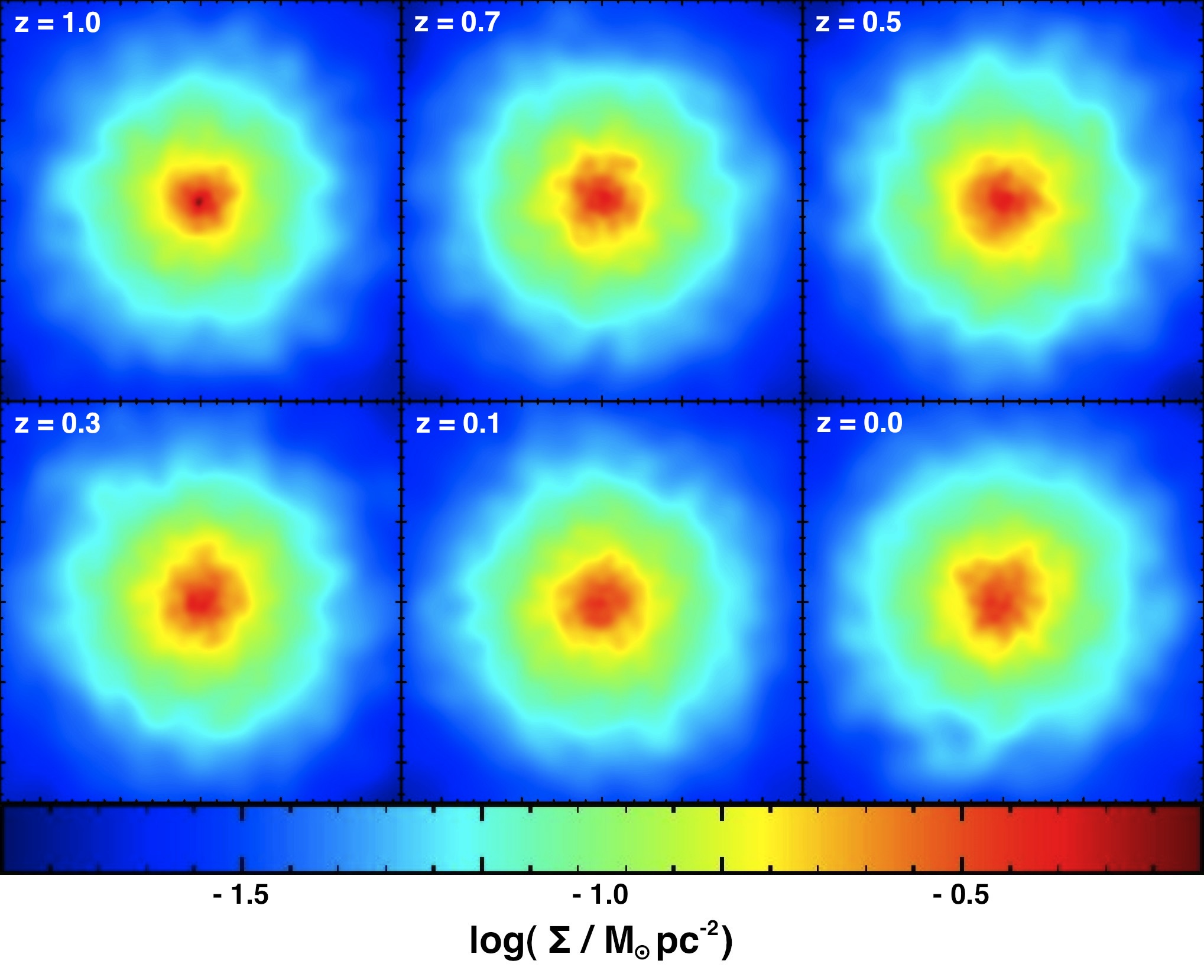}
\caption{Same as Figure \ref{fig:dms1mpc}, but for the core of the
  halo with a panel side length of 50 kpc. The envelope of the green
  region corresponds to approximately half the Hernquist scaleradius.}
\label{fig:dms50kpc}
\end{figure}

We finally need to check whether the Gaussian DM-spheres remain bound
when they enter the parent halo, or if they are quickly dissolved such
that they can be seen as smooth accretion as intended. For this we
compute surface density maps of the system at six redshifts. In Figure
\ref{fig:dms1mpc} the maps are plotted on the scale of 1 Mpc for the
model with the fiducial parameters. As the virial radius is of the
order of 200 kpc (depending on redshift), the total halo is shown. In
the first panel at $z=1$, all 400 DM-spheres are clearly visible, as
their density is larger than the surrounding background density. As
time elapses, the DM-spheres fall towards the centre of the halo. When
the DM-spheres enter the halo, they are able to stay bound for a short
amount of time, corresponding to less than half an orbital period, and
are then dissolved. In the last panel at $z=0$, most of the DM-spheres
have been destroyed. Only those DM-spheres that have just entered the
halo are still bound as they have not been within the halo long enough
to be tidally destroyed.

As most of the DM-spheres can remain bound for almost half an orbital
period, they are able to reach the pericentre of their orbit before
they are dissolved. If the DM-spheres are freely falling, this would
be problematic for a central disc of stars, since the constant
bombardment of small objects can disturb the disc and possibly thicken
it. Therefore we modelled the blobs with an initial velocity, such that
the pericentre is large enough and the DM-spheres do not perturb the
disc. In Figure \ref{fig:dms50kpc} the surface density maps are
plotted on a scale of 50 kpc such that the region within half of the
Hernquist scalelength of the halo is shown. This is the typical scale
up to which exponential discs with scalelengths of a few kpc can be
resolved. As we can see, the density profile remains the same
throughout the simulation, and no DM-spheres can be identified in this
region as here the density of a DM-sphere is lower than the density of
the halo.  This shows that our method, which models the smooth
accretion with small bound objects instead of a uniform distribution
of dark matter particles, does not lead to a perturbation of central
objects.

\subsection{Simulations of Semi-Analytic Merger Trees}
\label{trees}

In this section we describe how we use the SAMs to populate $N$-body
merger trees with galaxies and then use them as the initial conditions
for hydrodynamical multiple merger simulations.

\begin{figure}
\psfig{figure=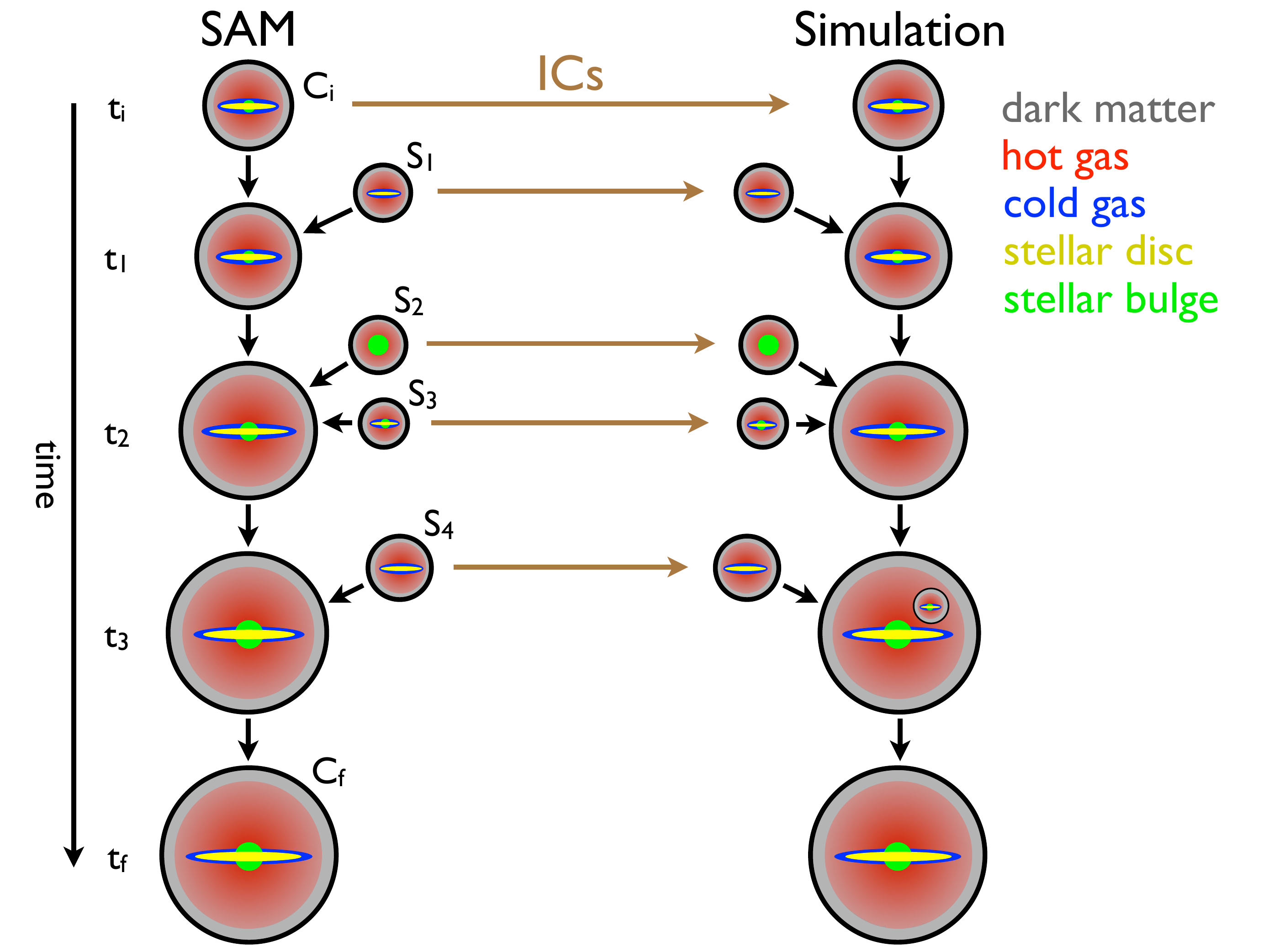,width=0.49\textwidth}
\caption{Schematic view of the combination of semi-analytic models and
  merger simulations. On the left side the semi-analytic merger tree
  is shown, with time running from top to bottom. At the starting time
  $t_i$, a numerical realization of the central galaxy is created
  using the properties of the central galaxy in the SAM at this
  time. This system is then simulated with the hydrodynamical code
  until the time of the first merger $t_1$, where the first satellite
  $S_1$ galaxy is put into the simulation using the properties
  predicted by the SAM. The resulting merger is simulated until the
  next galaxies ($S_2$ and $S_3$) enter the main halo at $t_2$, at
  which point they are also included in the simulation. This procedure
  is repeated for all mergers above a specified mass ratio until the
  final time of the run $t_f$.}
\label{fig:samtosim}
\end{figure}

A schematic view of our method is presented in Figure
\ref{fig:samtosim}. In a first step we select a dark matter merger
tree from the large-scale $N$-body simulation and use the SAM to
predict the properties of the baryonic components of each halo at
every timestep. A resulting galaxy merger tree is shown in the left
side of Figure \ref{fig:samtosim} with time running from top to
bottom. We then choose a starting time $t_i$ from which we want to
simulate this tree. In our simple example, the main system experiences
four mergers after $t_i$. We then use the predictions of the SAM for
the central galaxy of the main halo at $t_i$ and create a particle
realisation with the galaxy generator, as indicated by the brown
arrow. This model galaxy is shown in the top right of Figure
\ref{fig:samtosim} and in our example consists of a dark matter halo
(grey), a hot gaseous halo (red), a cold gaseous disc (blue), a
stellar disc (yellow) and a small stellar bulge (green). We evolve
this galaxy with our hydrodynamical code until the time $t_1$ when the
first satellite galaxy $S_1$ enters the main halo. A particle
realisation of the $S_1$ system (dark halo and galaxy) is then created
using the semi-analytic prediction for the galaxy properties and
included in the simulation at the virial radius of the main halo.  The
orbital parameters of $S_1$ (position and velocity at the time of
accretion) are taken directly from the $N$-body simulation. We evolve
this merger with the hydrodynamical code, until the next satellite
galaxy enters the main halo. In our example, two galaxies enter the
halo at the same time, and thus we create particle realisations using
the predictions of the SAM for both galaxies $S_2$ and $S_3$ and
include them in the simulation by positioning them at the virial
radius. This procedure is repeated for all merging satellites until
the final time $t_f$. In this way, we naturally include multiple
mergers, when an already merging galaxy has not been fully accreted as
the next galaxy is entering the halo. This is indicated at $t_3$, where
the galaxy $S_3$ is still orbiting while the satellite $S_4$ enters
the halo. In the example shown in Figure \ref{fig:samtosim}, the
central galaxy grows both through accretion of dark matter and gas
(e.g. from $t_i$ to $t_1$ or from $t_3$ to $t_f$) and through mergers
of satellite galaxies (e.g. from $t_1$ to $t_2$).

Our approach thus requires two main steps: creating particle
realisations of galaxies as predicted by the SAM and combining these
initial conditions in simulations as determined by the merger tree
(i.e. every galaxy has to enter the simulation at the specified time
and position).  This means we first have to specify how the
information about the galaxy properties that is computed with the SAM
is transformed into three dimensional particle based galaxy models
that can be simulated with a hydrodynamic code. Then we have to
determine how the satellite galaxies are included in the simulation,
so their position and velocity have to be calculated.

We note that the starting time $t_i$, or the starting redshift $z_i$,
respectively, can also be chosen such that the simulation starts at a
very early epoch. In this case the central galaxy would consist only
of a dark matter halo and hot gas in this halo. The dark matter will
then grow by mergers and smooth accretion, and the stellar disc and
bulge will form as a result of cooling and accretion of gas, and
merger events.

\subsubsection{Creating particle realisations of semi-analytic galaxies}

In order to create particle realisations of the semi-analytic
galaxies, we need to specify those galaxies that are included in our
simulation.  For this we first have to choose a starting redshift
$z_i$ and a minimum merger ratio $\mu_{\rm min}$. We define this ratio
as the mass of the dark matter in the entering subhalo divided by the
dark matter mass of the main halo at the time the satellite passes the
virial radius.

The starting redshift and the minimum merger ratio are in principle
free parameters and can be chosen according to the physical problem
one wants to address. This makes our approach very flexible.

In a semi-analytic merger tree we identify the central galaxy at the
starting redshift, all satellite galaxies within the virial radius at
this time (which have not merged with the central galaxy yet) and all
satellite galaxies that enter the main halo at a later time. From
these galaxies we only select those that fulfill our merger mass ratio
criterion. For every selected galaxy we record the value for the time
the galaxy enters the main halo $t_{\rm enter}$, the virial mass of
the dark matter halo $M_{\rm vir}$, its concentration $c$, its spin
parameter $\lambda$, and the semi-analytic predictions for the masses
of the hot gaseous halo $M_{\rm hg}$, the cold gaseous disc $M_{\rm
  cg}$, the stellar disc $M_{\rm disc,*}$ and the stellar bulge
$M_{\rm bulge}$, and the scalelength of the stellar disc $r_{\rm
  disc}$. All these quantities are taken at $t_i$ for the central
galaxy and satellites that are already with the main halo at $t_i$,
and at $t_{\rm enter}$ for all satellite galaxies that are entering
the main halo after $t_i$. Additionally, for all satellite galaxies we
record the dynamical friction time.

The remaining structural parameters of every galaxy that are not
directly predicted by the SAM are based on empirical scalings. The
scaleheight of the stellar disc is assumed to be a fraction of its
scalelength $z_0=\zeta r_{\rm disc}$, where typically
$\zeta=0.15$. Similarly the scalelength of the gaseous disc is related
to that of the stellar disc by $r_{\rm gas}=\chi r_{\rm disc}$, with a
standard value of $\chi=1.5$.  The core radius $r_c$ of the hot
gaseous halo is related to the scalelength of the dark matter halo
$r_s$ by $r_c=\xi r_s$ with a fiducial value of $\xi=0.22$ and the
slope parameter of the $\beta$-profile is set to $\beta_{\rm hg}=2/3$.

In order to select the number of particles in every component of the
galaxies, we determine the semi-analytic prediction for the final
stellar mass of the central galaxy $M_{*,f}$ and choose a number of
stellar particles in the central galaxy $N_*$ that we would like to
obtain at the end of our simulation in order to get the stellar
particle mass $m_*=M_{*,f}/N_*$.  As the simulation code produces
$N_g$ generation of stellar particles from every gas particle due to
star formation (where $N_g$ is typically 2), we set the mass of every
gas particle to $m_{\rm gas}=N_gm_*$. In this way, all stellar
particles (old and new) have the same mass. The mass of the dark
matter particles is finally selected as $m_{\rm dm}=\kappa m_*$ where
$\kappa$ is a free parameter. If $\kappa$ is chosen too low, the dark
matter particles have a very low mass which results in a very large
number of dark matter particles and thus in a high computational
cost. For high values of $\kappa$ the mass of the dark matter
particles can become too large, such that these massive particles
perturb the disc component which results in numerical disc
heating. Typically values of $\kappa\sim15$ are both computationally
efficient and lead to no measurable heating.
In order to prevent a component from being unstable due to a low
number of particles, we remove a component if its number of particles
is lower than a minimum particle threshold $N_{\rm min}$.

Additionally, for simulations where we are interested in the satellite
population and not in the central galaxy, we can allow for a higher
resolution in the satellite galaxies. This is done by dividing the
mass of the particles of every satellite by a number $N_{\rm
  res,sat}$. This number, however, should not be chosen to be too
high, since the particle masses of the satellite would be much lower
than the particle masses of the main system. This can lead to mass
segregation, which means that the particles of higher mass (i.e. the
central galaxy particles) will preferentially settle at the bottom of
the potential. We wish to avoid this numerical effect which requires
that $N_{\rm res,sat}\lta10$.

The last parameter that needs to be fixed is the gravitational
softening length $\epsilon$. In order to ensure that the maximum
gravitational force exerted from a particle is independent of its
mass, we scale the softening lengths of all particle species (dark
matter, gas and stars) with the square root of the particle mass
\citep{dehnen2001}. The normalisation of this relation is obtained
with a free parameter $\epsilon_1$ that specifies the softening length
for a particle of one internal mass unit of the code
(i.e. $10^{10}\Msun$). The softening length is thus given as $\epsilon
= \epsilon_1 \sqrt{m_{\rm part}/10^{10}\Msun}$, where $m_{\rm part}$ is
the mass of the particle. For our fiducial choice of
$\epsilon_1=32\kpc$ and a typical particle mass of $10^{5}\Msun$ this
results in a softening length of 100 pc.

Using this recipe, we create particle realisations of all selected
galaxies. These can now be used as initial conditions for the
hydrodynamical simulation. The next task is thus to specify when a
galaxy enters the main halo, and at which positions and with what
velocity.

\subsubsection{Performing the multiple merger simulation}

\begin{table*}
 \centering
 \begin{minipage}{140mm}
  \caption
  {Summary of the parameters used for the simulations of merger trees and their fiducial value.}
    \null
  \begin{tabular}{@{}llr@{}}
  \hline
  Parameter & Description & Fiducial value\\
 \hline
$z_i$ & Redshift at the start of the simulation & 1.0\\
$\mu_{\rm min}$ & Minimum dark matter mass ratio & 0.03\\
$\zeta$ & Ratio of scaleheight and scalelength of the stellar disc & 0.15\\
$\chi$ & Ratio of scalelengths between gaseous and stellar disc & 1.5\\
$\xi$ & Ratio of gaseous halo core radius and dark matter halo scaleradius & 0.22\\
$\beta_{\rm hg}$ & Slope parameter of gaseous halo & 0.67\\
$\alpha$ & Ratio of specific angular momentum between gaseous and dark halo& 4.0\\
$N_*$ & Expected final number of stellar particles in the central galaxy & $200\,000$\\
$\kappa$ & Ratio of dark matter and stellar particle mass & 15.0\\
$N_{\rm res,sat}$ & Ratio of satellite and central galaxy particle mass & 1.0\\
$N_{\rm min}$ & Minimum number of particles in one component & 100\\
$\epsilon_1$ & Softening length in kpc for particle of mass $m=10^{10}\Msun$ & 32.0\\
$t_0^*$ & Gas consumption time-scale in Gyr for star formation model & 3.5$^\dagger$\\
$A_0$ & Cloud evaporation parameter for star formation model & 1250.0$^\dagger $\\
$\beta_{\rm SF}$ & Mass fraction of massive stars for star formation model & 0.16$^\dagger $\\
$T_{\rm SN}$ & Effective supernova temperature in K for feedback model& $1.25\times10^{8\dagger}$\\
$\eta$ & Mass loading factor for wind model & 1.0\\
$v_{\rm wind}$ & Initial wind velocity in \kms for wind model & 500.0\\
\hline
\null\\
\multicolumn{3}{l}{$^\dagger $ The star formation parameters assume a Kroupa IMF.}\\
\label{t:smtparameters}
\end{tabular}
\end{minipage}
\end{table*}

We start with the particle realisation of the central galaxy and move
into its rest frame. For all satellite galaxies we have to compute the
relative position and velocity with respect to the central galaxy at
the time when they are included in the simulation. Since we use
merger trees drawn from simulations, we can directly extract the
relative initial positions and velocities from the tree. As $t_{\rm
  enter}$ is defined as the time when a satellite galaxy passes the
virial radius of the main halo, the initial distance is always equal
to the virial radius at $t_{\rm enter}$.

However, we have to divide between those galaxies that are already
within the main halo at $z_i$ and those galaxies that enter the halo
at a later time. For those galaxies that have entered the halo before
$z_i$, we expect that they have already lost some of their angular
momentum due to dynamical friction and are therefore closer to the
central galaxy than the virial radius. Therefore we scale their
initial distance and velocity by
\begin{equation}
r^\prime = r \sqrt{1-\frac{t_i-t_{\rm enter}}{t_{\rm df}}} \quad {\rm and} \quad v^\prime = v \sqrt{1-\frac{t_i-t_{\rm enter}}{t_{\rm df}}}\,,
\end{equation}
where $t_i$ is the starting time of the simulation, $t_{\rm enter}$ is the time the satellite entered the main halo and $t_{\rm df}$ is the
dynamical friction time (i.e. the time it takes the satellite to merge with the central galaxy). The direction of the position and velocity vectors
are unaltered.

Before the simulation is started we include all satellite galaxies
that have entered the main halo before $t_i$ and all satellites that
are entering the halo at $t_i$ in the initial conditions. We evolve
this system with the hydrodynamical code until the time when the first
satellite enters the main halo $t_{{\rm enter}, 1}$ as specified by
the merger tree. At this time we interrupt the evolution of the
simulation and include the particle realisation of this satellite
galaxy in the simulation. The simulation is then resumed and this
process is repeated until the final time of the simulation $t_f$.

The main advantage of including the satellite galaxies only when they
enter the main halo is that the computational cost is reduced. The
satellites are only simulated when they affect (or are affected by)
the main system. Up to this point we use the SAM to compute the
evolution of the satellite galaxy progenitor's properties.  In Table
\ref{t:smtparameters}, we summarise the free parameters of our model
and present our fiducial values. We have tested these values mainly on
merger trees of MW-like galaxies. For systems of higher or lower halo
mass they may have to be adjusted accordingly.  We also list the free
parameters that are used in the hydrodynamical code and its cooling,
star formation and wind models.

\section{Simulations of Milky Way-like galaxies}
\label{sec:sims}

\begin{table*}
 \centering
  \begin{minipage}{140mm}
  \caption
  {Properties of the galaxies in the merger trees. For the main system the properties are given at the starting time of the simulation $t_i$, while
  for the satellites the properties are given at the time when they enter the main halo $t_{\rm enter}$. All masses are in solar units and all scales
  are given in kpc.}
    \null
  \begin{tabular}{@{}lrrrrrrrrr@{}}
  \hline
ID & $t_{\rm enter}$ & $t_{\rm merge}~^a$ & ~$\mu^{-1}$ & $\log(M_h)$ & ~~~~~$c$ & $\log(M_*)$ & ~$B/T$ & ~$f_{\rm gas}~^b$ & ~$r_{\rm disc}$\\
\hline
\hline
\multicolumn{10}{c}{Tree 1775}\\
\hline
Main & 6.25 & - & - & 11.55 & 4.71 & 10.16 & 0.15 & 0.41 & 2.53\\
Sat 1 & 7.72 & 14.47 & 14.42 & 10.43 & 7.01 & 7.47 & 0.00 & 0.90 & 1.13\\
Sat 2 & 7.88 & 13.33 & 9.48 & 10.68 & 6.77 & 9.25 & 0.37 & 0.37 & 1.36\\
Sat 3 & 11.76 & 15.47 & 8.04 & 10.96 & 9.01 & 9.22 & 0.00 & 0.43 & 1.87\\
\hline
\hline
\multicolumn{10}{c}{Tree 1808}\\
\hline
Main & 6.25 & - & - & 11.70 & 4.56 & 10.11 & 0.00 & 0.51 & 1.65\\
Sat 1 & 6.90 & 13.48 & 9.87 & 10.76 & 6.02 & 8.92 & 0.00 & 0.37 & 1.04\\
Sat 2 & 7.07 & 12.52 & 15.58 & 10.62 & 6.30 & 8.73 & 0.00 & 0.58 & 1.66\\
Sat 3 & 10.42 & 22.29 & 22.80 & 10.57 & 8.71 & 8.66 & 0.18 & 0.53 & 1.43\\
\hline
\hline
\multicolumn{10}{c}{Tree 1877}\\
\hline
Main & 6.25 & - & - & 11.53 & 4.73 & 9.88 & 0.00 & 0.41 & 2.08\\
Sat 1 & 6.25 & 10.03 & 7.80 & 10.64 & 5.73 & 8.81 & 0.00 & 0.56 & 1.06\\
Sat 2 & 6.74 & 14.16 & 13.12 & 10.53 & 6.17 & 8.78 & 0.00 & 0.41 & 0.91\\
Sat 3 & 9.49 & 24.84 & 16.86 & 10.51 & 8.14 & 8.17 & 0.00 & 0.71 & 1.46\\
Sat 4 & 9.64 & 23.02 & 10.57 & 10.75 & 7.91 & 8.76 & 0.00 & 0.70 & 2.69\\
Sat 5 & 9.96 & 15.56 & 8.77 & 10.93 & 7.83 & 9.26 & 0.05 & 0.56 & 2.53\\
Sat 6 & 13.31 & 27.59 & 23.16 & 10.58 & 10.95 & 6.41 & 0.00 & 1.00 & 4.19\\
\hline
\hline
\multicolumn{10}{c}{Tree 1968}\\
\hline
Main & 6.25 & - & - & 11.60 & 4.66 & 10.02 & 0.00 & 0.41 & 3.09\\
Sat 1 & 6.08 & 13.76 & 10.52 & 10.52 & 5.72 & 8.22 & 0.00 & 0.70 & 1.38\\
Sat 2 & 6.41 & 11.04 & 6.08 & 10.86 & 5.61 & 8.80 & 0.00 & 0.74 & 1.96\\
Sat 3 & 6.57 & 20.96 & 21.52 & 10.41 & 6.21 & 7.96 & 0.00 & 0.81 & 1.65\\
Sat 4 & 12.05 & 39.86 & 21.00 & 10.57 & 9.96 & 8.57 & 0.06 & 0.70 & 2.13\\
\hline
\hline
\multicolumn{10}{c}{Tree 1975}\\
\hline
Main & 6.25 & - & - & 11.57 & 4.69 & 9.84 & 0.05 & 0.68 & 3.57\\
Sat 1 & 5.27 & 8.31 & 6.08 & 10.61 & 5.07 & 8.62 & 0.16 & 0.61 & 0.95\\
Sat 2 & 7.39 & 14.30 & 24.23 & 10.24 & 7.05 & 7.70 & 0.00 & 0.82 & 0.92\\
Sat 3 & 7.56 & 12.81 & 5.42 & 10.97 & 6.18 & 9.33 & 0.09 & 0.52 & 2.50\\
Sat 4 & 7.56 & 18.16 & 10.87 & 10.66 & 6.56 & 8.11 & 0.24 & 0.92 & 2.30\\
Sat 5 & 8.37 & 22.92 & 26.00 & 10.46 & 7.41 & 8.11 & 0.08 & 0.79 & 1.52\\
Sat 6 & 8.85 & 27.09 & 21.83 & 10.58 & 7.57 & 8.31 & 0.00 & 0.82 & 2.23\\
\hline
\hline
\multicolumn{10}{c}{Tree 1990}\\
\hline
Main & 6.25 & - & - & 11.37 & 4.90 & 9.66 & 0.00 & 0.53 & 2.15\\
Sat 1 & 6.41 & 11.35 & 6.81 & 10.55 & 5.94 & 8.28 & 0.00 & 0.74 & 1.55\\
Sat 2 & 6.74 & 13.44 & 10.13 & 10.50 & 6.22 & 8.17 & 0.04 & 0.70 & 0.84\\
Sat 3 & 9.80 & 18.33 & 14.04 & 10.63 & 8.18 & 8.72 & 0.17 & 0.78 & 2.87\\
Sat 4 & 10.11 & 15.01 & 10.39 & 10.83 & 8.10 & 9.07 & 0.00 & 0.30 & 1.45\\
\hline
\null\\
\multicolumn{10}{l}{$^a$ Time when satellite merges onto central galaxy, computed with dynamical friction recipe}\\
\multicolumn{10}{l}{$^b$ Gas fraction in the disc}\\
\label{t:mergertrees}
\end{tabular}
\end{minipage}
\end{table*}

\begin{table*}
 \begin{minipage}{140mm}
{\bf Table \ref{t:mergertrees}} cont.\\
    \null\\
  \begin{tabular}{@{}lrrrrrrrrr@{}}
  \hline
ID & $t_{\rm enter}$ & $t_{\rm merge}~~$ & ~$\mu^{-1}$ & $\log(M_h)$ & ~~~~~$c$ & $\log(M_*)$ & ~$B/T$ & ~$f_{\rm gas}~~$ & ~$r_{\rm disc}$\\
\hline
\hline
\multicolumn{10}{c}{Tree 2048}\\
\hline
Main & 6.25 & - & - & 11.50 & 4.76 & 9.97 & 0.01 & 0.36 & 1.15\\
Sat 1 & 6.57 & 11.26 & 9.81 & 10.58 & 6.03 & 8.48 & 0.00 & 0.86 & 3.37\\
Sat 2 & 6.90 & 11.80 & 8.49 & 10.78 & 6.01 & 8.82 & 0.00 & 0.67 & 2.39\\
Sat 3 & 7.23 & 15.89 & 20.99 & 10.51 & 6.59 & 9.35 & 0.03 & 0.00 & 1.07\\
\hline
\hline
\multicolumn{10}{c}{Tree 2092}\\
\hline
Main & 6.25 & - & - & 11.39 & 4.88 & 9.79 & 0.24 & 0.56 & 2.15\\
Sat 1 & 6.08 & 10.61 & 14.27 & 10.20 & 6.07 & 7.84 & 0.00 & 0.83 & 1.66\\
Sat 2 & 6.25 & 13.08 & 11.11 & 10.34 & 6.04 & 8.28 & 0.00 & 0.78 & 1.54\\
Sat 3 & 8.37 & 20.41 & 11.29 & 10.68 & 7.12 & 9.01 & 0.00 & 0.56 & 2.76\\
\hline
\hline
\multicolumn{10}{c}{Tree 2126}\\
\hline
Main & 6.25 & - & - & 11.52 & 4.75 & 9.87 & 0.05 & 0.38 & 1.12\\
Sat 1 & 7.23 & 11.65 & 8.07 & 10.69 & 6.32 & 8.84 & 0.00 & 0.42 & 1.14\\
Sat 2 & 7.56 & 14.06 & 11.33 & 10.67 & 6.57 & 8.94 & 0.00 & 0.40 & 1.55\\
Sat 3 & 11.91 & 17.82 & 8.30 & 10.90 & 9.22 & 9.34 & 0.30 & 0.55 & 2.60\\
Sat 4 & 12.20 & 22.71 & 14.57 & 10.72 & 9.79 & 8.74 & 0.00 & 0.67 & 2.36\\
Sat 5 & 13.45 & 23.69 & 21.42 & 10.63 & 10.97 & 7.97 & 0.00 & 0.94 & 3.38\\
\hline
\hline
\multicolumn{10}{c}{Tree 2181}\\
\hline
Main & 6.25 & - & - & 11.46 & 4.80 & 9.66 & 0.00 & 0.70 & 3.90\\
Sat 1 & 7.56 & 10.64 & 6.62 & 10.70 & 6.53 & 8.79 & 0.00 & 0.44 & 1.25\\
Sat 2 & 8.37 & 11.02 & 13.31 & 10.54 & 7.32 & 7.07 & 0.00 & 0.98 & 1.79\\
Sat 3 & 8.37 & 19.33 & 20.04 & 10.36 & 7.56 & 8.23 & 0.00 & 0.72 & 1.54\\
Sat 4 & 8.53 & 12.18 & 9.71 & 10.74 & 7.15 & 8.87 & 0.00 & 0.49 & 1.44\\
Sat 5 & 8.85 & 17.34 & 10.50 & 10.81 & 7.24 & 9.50 & 0.03 & 0.21 & 1.38\\
\hline
\end{tabular}
\end{minipage}
\end{table*}

We employ our model in a series of simulations of MW-like galaxies,
i.e. disc galaxy systems with a dark matter halo mass of
$M_h\approx10^{12}\Msun$.  The aim of this study is to investigate how
a central disc galaxy evolves from redshift $z=1$ to $z=0$ while
experiencing minor mergers.

We use dark matter merger trees drawn from the $N$-body simulation
presented in section \ref{sec:mwmergers}. As we are interested in
systems that experience only minor mergers after $z=1$, we select 10
trees with a final halo mass of $M_h\approx10^{12}\Msun$, that have
only mergers with a mass ratio of $\mu<0.2$. The properties of these
trees are presented in Table \ref{t:mergertrees}. As a starting
redshift we use $z_i=1$ which corresponds to a cosmic time of
$t=6.25\Gyr$. We employ the fiducial model parameters presented in
Table \ref{t:smtparameters} and simulate all trees up to a redshift
$z=0$.

\begin{figure*}
\centering
\includegraphics[width=0.92\textwidth]{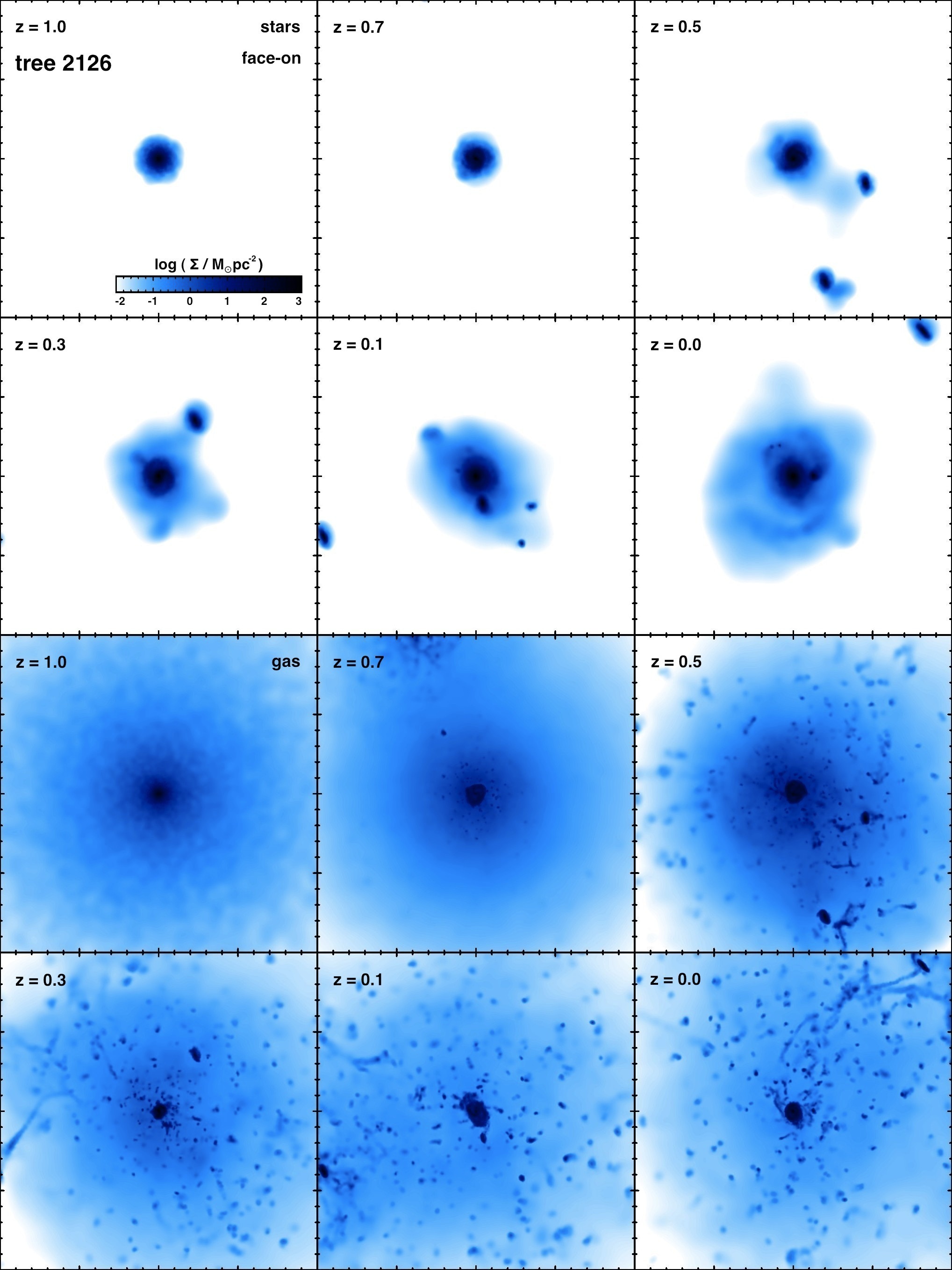}
\caption{Projected surface density for the stellar component (upper
  two rows) and the gaseous component (lower two rows) for the typical
  merger tree \textit{2126} as viewed from face-on. Each panel
  measures 200 kpc on a side and the redshift is displayed in the
  upper left corner of each panel.}
\label{fig:2126face}
\end{figure*}

\begin{figure*}
\includegraphics[width=0.92\textwidth]{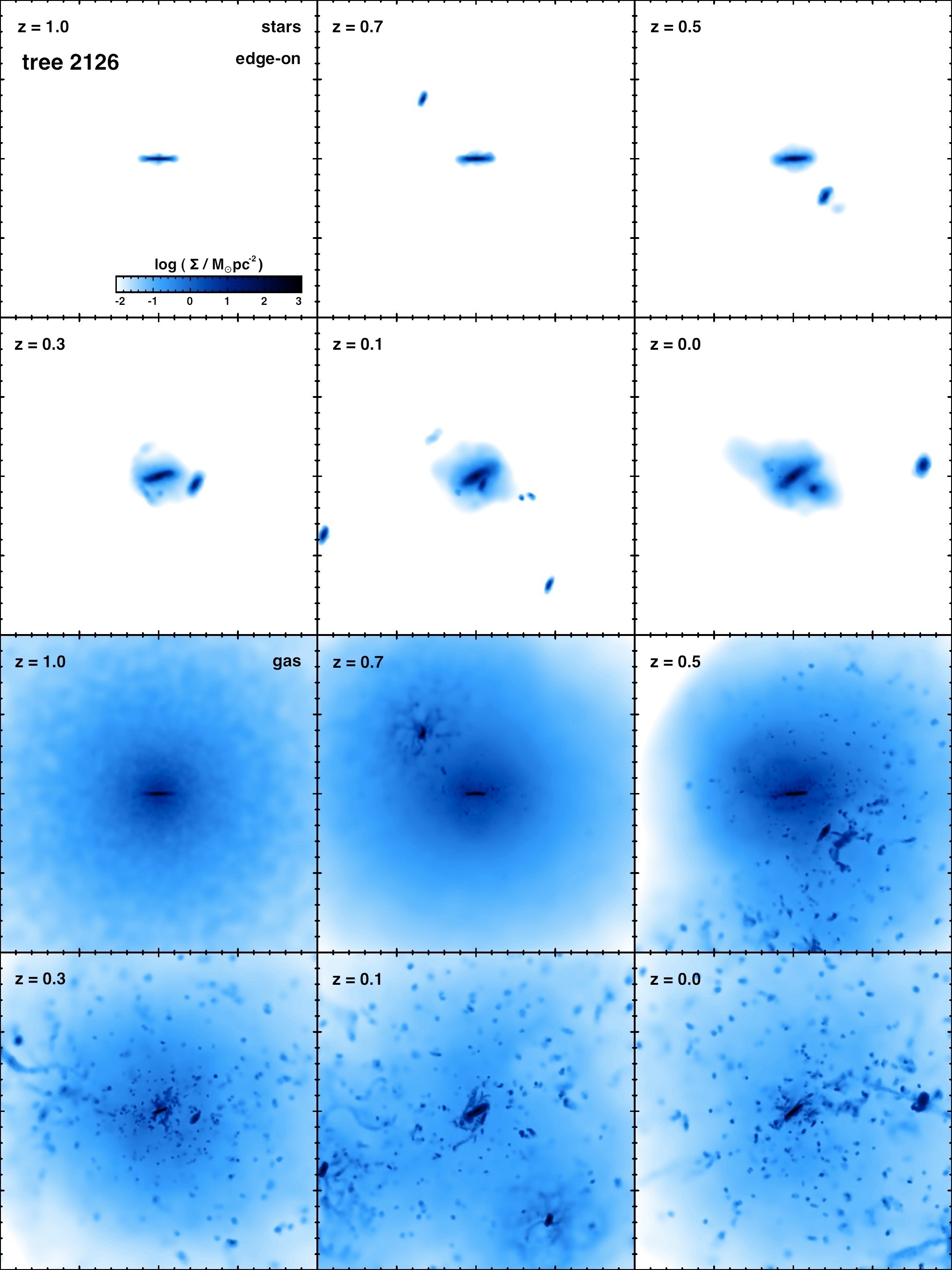}
\caption{Similar to Figure \ref{fig:2126face}, but for an edge-on projection.}
\label{fig:2126edge}
\end{figure*}

The surface density for a typical merger tree (\textit{2126}) is shown
in Figures \ref{fig:2126face} (face-on projection) and
\ref{fig:2126edge} (edge-on projection) for the stellar component
(upper panels) and the gaseous component (lower panels). The central
galaxy evolves in isolation for $\sim1\Gyr$ until the first two
satellites enter the halo. These satellites pass the central galaxy at
a pericentric distance of 55 and 90 kpc, respectively. While the first
satellite continues orbiting around the central galaxy and
subsequently loses angular momentum and energy, which leads to a
decrease of the pericentric distance, the second satellite has enough
orbital energy to leave the main halo again. Near the end of the
simulation the last three satellites enter the halo and pass the
central galaxy.

During the evolution of the simulation, the stellar disc of the
central galaxy is heated and thickened.  At the same time, a stellar
halo forms as a result of two processes. First, due to the close
passage of the satellites, some stars in the disc are scattered out of
the disc, as energy and angular momentum is transferred. Second, some
satellite stars are accreted onto the central galaxy. These stars
still retain some of the angular momentum of the satellite such that
they do not settle in the disc, but remain in the halo. Additionally,
we can identify several stellar streams which originate from the
destruction of satellites. However, due to the limited resolution in
this pilot study, the number of particles in these streams is not very
high, so that a detailed study is not possible. This limitation can be
overcome in future studies by increasing the parameter $N_{\rm
  res,sat}$, so that there are more particles in the satellites and
thus more particles in the streams.
When satellite galaxies enter the halo, a fraction of their
cold gas is stripped and eventually can fall towards the centre of the
system. Thus some of the stars in the central galaxy form from gas
that has been stripped from satellites and accreted onto the central
gaseous disc.

\subsection{Evolution of the stellar mass}

In this pilot study, we analyse the properties of the central galaxy
in each merger tree and these galaxies' evolution during the course of
the simulation.  Clearly, the stellar mass of the central galaxy is an
important quantity. In order to measure the stellar mass in the
simulation we identify FOF groups for every snapshot with a linking
parameter of $b=0.2$. Each FOF group is then split with the {\sc
  Subfind} code into a set of disjoint subhaloes that contain all
particles bound to a local overdensity. The main halo which contains
the central galaxy is the most massive subhalo of the FOF group. The
stellar mass of the central galaxy is the total mass of all stellar
particles that belong to the main halo.

\begin{figure*}
\psfig{figure=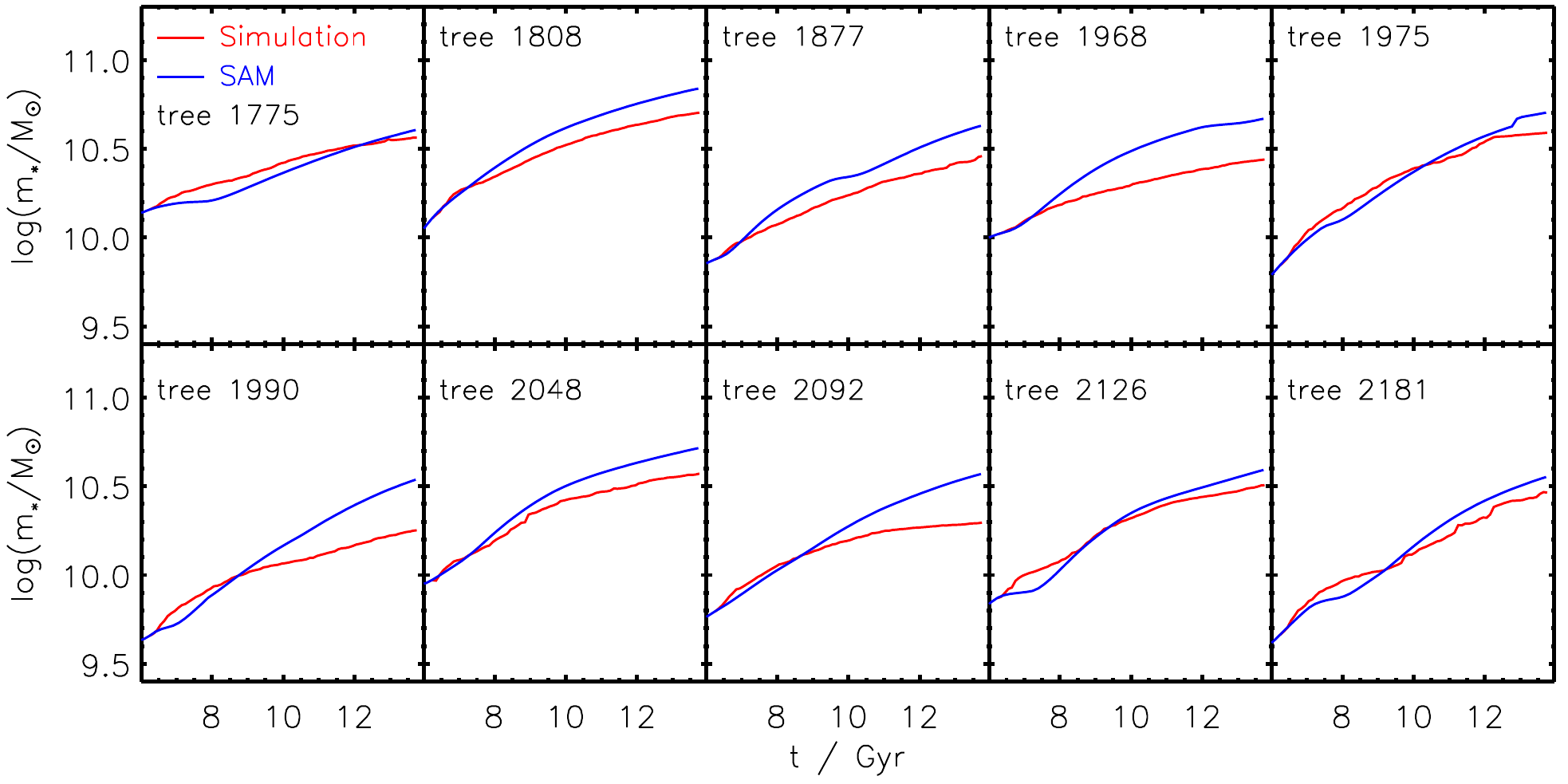,width=0.95\textwidth}
\caption{Evolution of the stellar mass in the central galaxy in each
  merger tree. The measured stellar mass in the simulation (red line)
  is compared to the prediction by the SAM (blue line).}
\label{fig:smass}
\end{figure*}

We plot the resulting evolution of the stellar mass for all merger
trees in Figure \ref{fig:smass}. The results of the simulations are
given by the red line, while the semi-analytic prediction is given by
the blue line. We find that all systems increase their stellar mass
and have a final stellar mass that is more than twice the value at
$z=1$. The stellar masses found in the simulation at $z=0$ range from
$\log(m_*/\Msun) = 10.3$ to $10.7$, while the majority of systems have
a stellar mass close to $\log(m_*/\Msun) = 10.5$. These results are in
excellent agreement with those obtained by \citet{moster2010a} using a
statistical halo occupation model. For some merger trees the
prediction by the SAM agrees very well with the result of the
simulation (e.g. the trees \textit{1775}, \textit{1975} and
\textit{2126}), while for other trees the semi-analytic prediction for
the final stellar mass is much higher than what we find in the merger
simulation (e.g. the trees \textit{1968}, \textit{1990} and
\textit{2048}). We note that the semi-analytic values for the stellar
mass tend to exceed, by a small amount, the average observed stellar
mass for a halo of mass $10^{12}\Msun$.
This is because the version of the SAM that we used had been optimized
for use with different merger trees, and we did not re-tune the
parameters to specifically match this constraint.

In the evolution of the stellar mass, we see that at some times the
mass increases quickly in a short time interval. This is not due to an
increased SFR due to an interaction-triggered starburst, but is a
result of the accretion of the satellite itself. The mass of the
accreted satellite is much larger than the newly formed mass as a
result of a starburst. Examples can be seen for tree
\textit{1975} ($t=11.5\Gyr$), tree \textit{2048} ($t=9\Gyr$) and tree
\textit{2181} ($t=9.6, 11.2$ and $12.1\Gyr$). Overall however, we find
that most stars in the central galaxy have formed from the cold
gaseous disc that is accreted from the gaseous halo (in-situ
formation), and a only few stars in the central galaxy originate from
accreted satellites (ex-situ formation).

\subsection{Evolution of the cold gas fraction}

\begin{figure*}
\psfig{figure=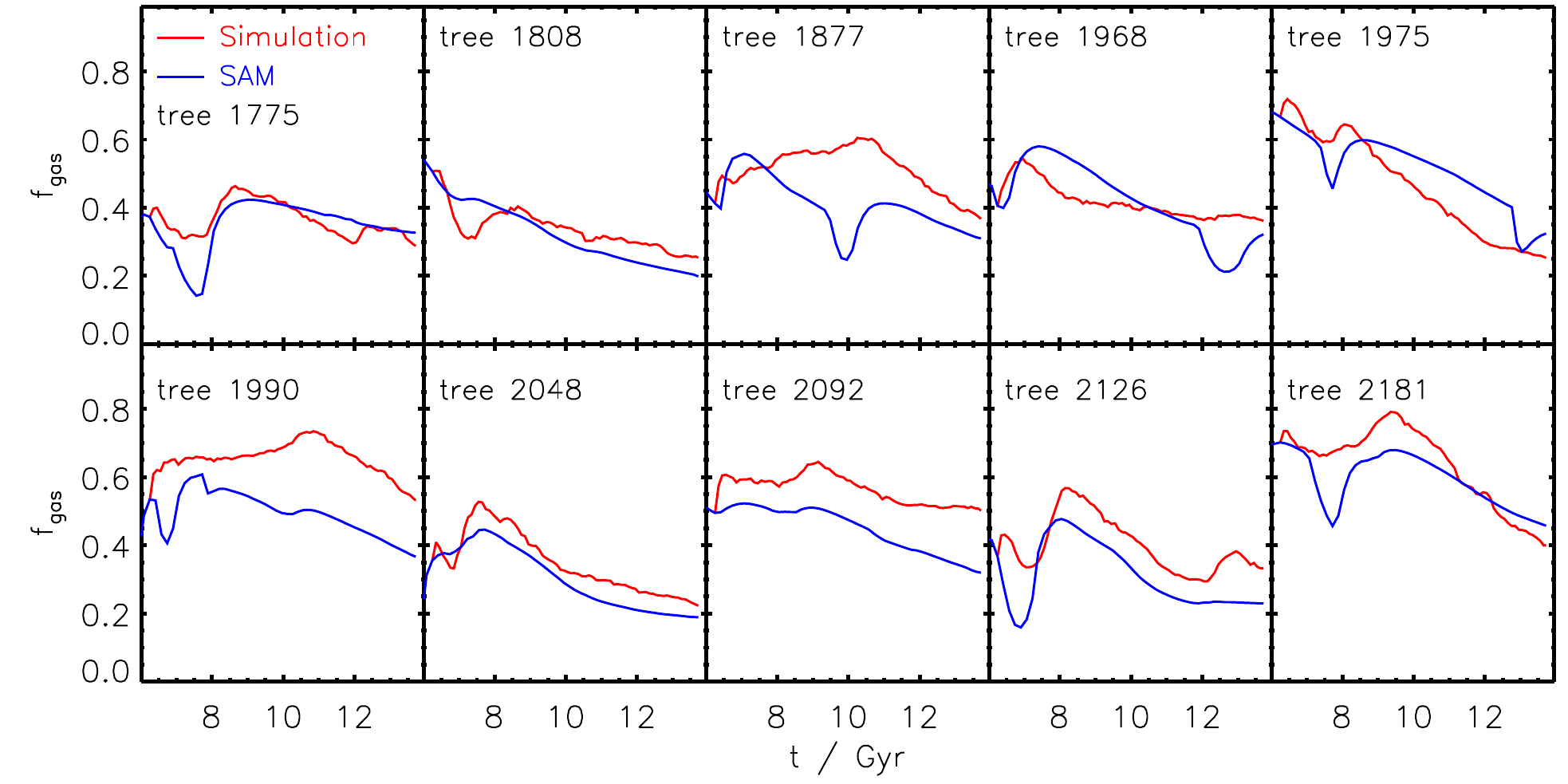,width=0.95\textwidth}
\caption{Cold gas fraction of the central galaxy in each merger
  tree. The cold gas fraction in the simulation is given by the red
  line, while the blue line shows the cold gas fraction in the SAM.}
\label{fig:fgas}
\end{figure*}

We further study the evolution of the cold gas component, i.e. all gas
particles with a temperature below $T=2\times10^4K$ that reside in the
central disc. The resulting gas fractions are plotted in Figure
\ref{fig:fgas}. We find that all systems have a higher gas fraction at
$z=1$, with typical values of 30 to 50 per cent. Towards $z=0$ the gas
fraction then decreases to lower values of 20 to 40 per cent. These
results are in very good agreement with those obtained by
\citet{stewart2009} based on indirect gas fraction estimates from star
formation rate densities.

In all systems that have just accreted a subhalo, the gas fraction is
elevated. This happens during the first passage of the satellite, in
which a considerable amount of its cold gas is unbound by tidal stripping. 
This gas then falls onto the central galaxy, leading to an increased
SFR. We note that unlike in the SAM, this usually happens during the
first passage, and not in the final coalescence, which can be much
later.  The semi-analytic predictions for the gas fractions agree very
well with those that are found in the simulations. This is due to the
very similar cooling rates in the halo for the SAM and the
simulations. The accretion of cooled gas from the halo is much larger
than the accretion of stripped cold gas from satellite
galaxies. Although the amount of accreted gas from satellites is
different in the SAM and the simulations, this effect is small, so
that the total gas fractions in the central discs agree quite well.

\subsection{Evolution of the scale length and height}

\begin{figure*}
\psfig{figure=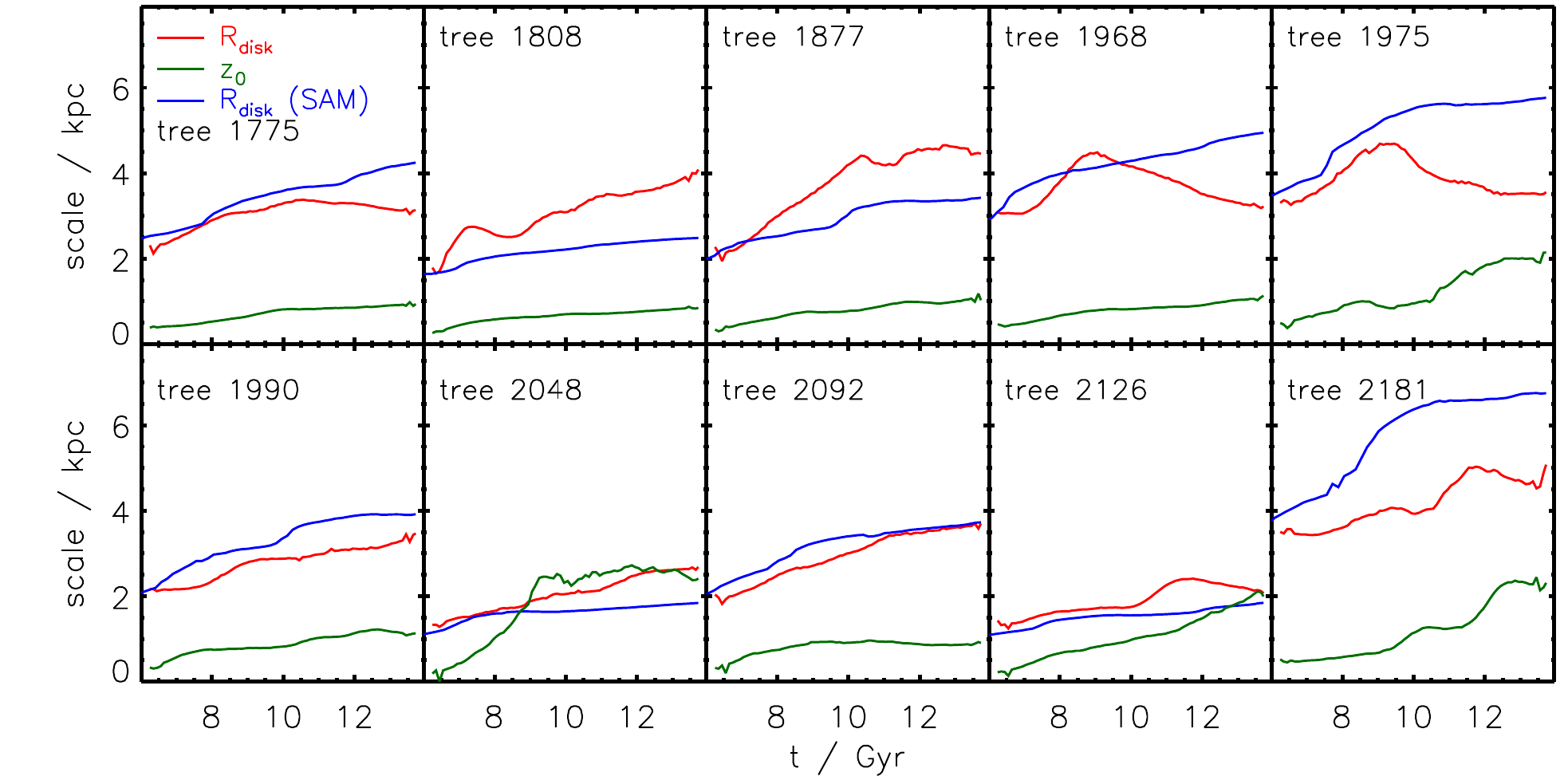,width=0.95\textwidth}
\caption{Scale parameters of the central galaxy in each merger
  tree. The red and green lines show the measured disc scalelengths
  and scaleheights, respectively.  The prediction for the disc
  scalelength by the SAM is added for comparison (blue line). }
\label{fig:scalelength}
\end{figure*}

Finally we study the evolution of the scale parameters of the stellar
components, i.e. the scalelength and height of the stellar disc.  In
order to measure these quantities in our simulations, we first use the
decomposition of the particles into disc and spheroidal components.
For the disc component we then compute the face-on projected surface
density profile and fit the exponential disc scalelength. Similarly,
we compute the projected edge-on surface density at a radius of
$R=8\kpc$, and fit a ${\rm sech}^2$ function to it to obtain the disc
scaleheight.

We plot the resulting scale parameters in Figure \ref{fig:scalelength}
for all merger trees. The disc scalelength $R_{\rm disc}$ is given by
the red line and the disc scaleheight $z_0$ is given by the green
line.  For comparison we included the prediction for the disc
scalelength by the SAM which is shown by the blue line. For some
systems, the scalelength in the simulation and the SAM agree very well
(trees \textit{1775}, \textit{1990} and \textit{2092}), while for
other systems the scalelength predicted by the SAM differs from the
one found in the simulation (trees \textit{1808}, \textit{1877} and
\textit{1975}). Interestingly, while the scalelength in the SAM always
increases with time, the scalelength measured in the simulation can
both increase and decrease.  Overall, we find a broad range of
scalelengths for MW-like systems, consistent with observations
\citep{barden2005}.

The disc scaleheight evolves slowly with time, as long as no satellite
merges with the central galaxy. For most systems the ratio between the
scaleheight and scalelength is roughly constant throughout the
simulation. Final scaleheights range from $z_0=0.6$ to $1.2\kpc$,
consistent with results form observations of MW-like systems
\citep{schwarzkopf2000,yoachim2006}. In some merger trees (trees
\textit{1975} and \textit{2048}), the thin disc is completely
destroyed due to mergers and only a thick disc with a final
scaleheight of $\approx4\kpc$ remains.

\section{Discussion and Outlook}
\label{sec:discussion}

In order to study the connection between the internal structure of
galaxies and their formation histories and large scale environment, a
galaxy formation model must have high resolution, include gas physics,
and include the cosmological background.  The purpose of this paper is
to present a novel approach to study the evolution of galaxies by
combining semi-analytic models with numerical hydrodynamic merger
simulations. Using the predictions of the semi-analytic model as the
initial conditions for multiple merger simulations, we were able to
achieve high resolution at a fraction of the computational cost of
standard cosmological hydrodynamic methods. 

Hydrodynamical simulations of binary galaxy mergers have been used
extensively to study the evolution and morphological transformation of
single galaxies due to merger effects, although additional information
must be used to place them in a cosmological context. They can be
regarded as a method to study the effects of a single encounter in a
detailed manner, rather than the complete evolution of a galaxy. For
these merger simulations one creates pre-formed galaxies, that are
stable in isolation, with parameters drawn from a grid or motivated by
observations. These model galaxies are then set on an orbit and
evolved with a hydrodynamical code.

Although this technique is useful for studying the evolution of a
galaxy during a single encounter, it is not able to predict the
typical galaxy properties at a given redshift, as the cosmological
background, i.e. the merger history of each galaxy is not taken into
account. For this reason, we studied the merger histories of galaxies
using merger trees drawn from a large cosmological $N$-body
simulation.  We studied whether most mergers are binary or whether
they usually involve multiple galaxies, i.e. if two galaxies generally
have enough time to merge after their haloes have merged, before the
remnant merges with another galaxy. We found that the probability for
two mergers happening within a few halo dynamical times is always
higher than the probability for many dynamical times to elapse
between mergers. In our merger trees, for more than 50 per cent of all
merger pairs, the second satellite enters the halo within three
dynamical times after the first satellite entered, independent of mass
ratio. This indicates that multiple mergers are more common than
sequences of isolated binary mergers. As a consequence, it is not
sensible to focus on binary mergers, but rather on merger simulations
which consider several satellites that enter the parent halo one after
each other.

Due to the large number of parameters involved in mergers (orbit,
masses, gas fractions, merger ratios, etc.)  it is impossible to cover
the whole parameter space in merger simulations by drawing the
parameters uniformly from a grid. Instead, a simple and elegant path
is to use semi-analytic galaxy merger trees to generate the initial
conditions for galaxy merger simulations. In this way, we
automatically select mergers that are expected to be common in the
Universe.  However, in order to do this, we had to extend the code
that creates the particle representations of galaxies used in
simulations, as smooth accretion of dark matter and the hot gaseous
halo component were not taken into account before. In order to model
the cooling and accretion of gas onto the disc we included a slowly
rotating hot gaseous halo in the initial conditions generator. We
modelled the smooth accretion of dark matter material that is too
small to be resolved as a halo in the merger trees, by placing
additional dark matter particles around the halo. These particles were
placed into small spherical systems with a Gaussian density profile to
represent the many sub-resolution systems that are expected to be
accreted. The distance to the halo centre was chosen such that they
fall into the virial radius of the halo at a specified time extracted
from the merger tree. In order to model a system that is stable in
isolation, we computed the distribution function for a Gaussian
profile and the velocity of each particle with a rejection sampling
technique. We tested this smooth accretion model for an isolated
MW-sized halo and found an excellent agreement with the results from a
full cosmological simulation. Furthermore we verified that the small
spherical systems are quickly dissolved once they enter the halo.

With the extended model for the initial galaxies, we were able to
develop our novel approach that uses semi-analytic predictions as
initial conditions for a multiple merger simulation. Choosing a
starting redshift, we first created particle representations of the
central galaxy at this time, and the satellite galaxies at the time
when they enter the main halo. We set the mass resolution by requiring
that the final number of stellar particles equals a fixed parameter,
provided the final stellar mass in the SAM and the simulation are
equal.  The multiple merger simulation was then performed by evolving
the central galaxy until the first satellite enters the main halo, at
which point the particle realisation of the satellite was included in
the simulation at the virial radius. After this the simulation was
resumed and the procedure repeated until the end of the run. We
applied our method for ten merger trees drawn from a large
dissipationless N-body simulation, for which the main systems are
MW-sized at $z=0$ and have no major merger after the starting redshift
$z=1$.  For all ten trees we analysed the evolution of the properties
of the central galaxy and found good agreement between the stellar
mass in our simulations at $z=0$ and observational
constraints. Similarly, the disc scale lengths in our simulations
agree well with the observed range of scale lengths. Overall we also
found good agreement with the predictions by the SAM.

There are many differences in the details of the star formation and
stellar feedback prescriptions in the hydrodynamic simulations and the
SAMs, and the values of the parameters controlling the subgrid
physics, and we made no attempt to tune these parameters to match. The
fact that there is none-the-less reasonable agreement between the
important parameters of the galaxies indicates that the two tools can
be used together despite the differences in the details. In the
future, it would be straightforward to modify the physical ingredients
of the SAM and/or hydro simulations to achieve even better agreement. 


Although we demonstrated the basic elements of our new approach here,
there are still a number of improvements that can be made and physical
processes that can be included. For example, cosmological hydrodynamic
simulations have shown that filamentary large scale structures can
efficiently feed cold gas into massive halos via ``cold streams''
\citep{keres2005,dekel2006}.  Because our method assumes accretion
from a spherically symmetric halo, it does not include this accretion
mode. This process is dominant in massive haloes at a redshift of
$z>2$. Thus one possible improvement of our method is to model these
cold streams.

Another improvement is the inclusion of additional or more realistic
feedback processes. In our simulations we have implemented winds with
constant mass loading factor and constant wind velocities for all
galaxies. It has been shown that better agreement with observations
can be obtained with more sophisticated wind models, in which the mass
loading factor and wind velocity scales with galaxy properties,
e.g. momentum-driven winds \citep{oppenheimer2006}. More detailed
chemical evolution recipes that follow multiple elements can be
implemented to track abundance ratios in various galactic components.
Another form of feedback that is not considered in our simulations yet
is AGN feedback. The processes that result from gas accretion onto a
central BH can provide a large amount of energy, which may be able to
reduce cooling from the hot halo after a major merger.

We plan to use the results of simulations carried out with our new
approach to study the efficiency of starbursts and morphological
transformation (formation of spheroids) in mergers. This will lead to
more physical recipes that can then be fed back into the SAM.

In conclusion, we anticipate that the approach developed here will open a
path to many possible future applications, including (but not limited
to) the study of the formation of stellar halos through the accretion
and stripping of satellites, sub-structure and streams in galaxy
halos, the build-up of spheroids and thickening of galactic discs
through mergers, and the evolution of spheroid sizes and internal
velocities through mergers.

\section*{Acknowledgements} 

We thank Hans-Walter Rix and Kathryn Johnston, 
for enlightening discussions and useful comments on this work.
The numerical simulations used in this work were performed
on the THEO cluster of the Max-Planck-Institut f\"ur Astronomie
at the Rechenzentrum in Garching.
AVM acknowledges support from the  Sonderforschungsbereich
SFB 881 ``The Milky Way System'' (sub-project A1) 
of the German Research Foundation (DFG).


\bibliographystyle{mn2e}
\bibliography{moster2012}

\label{lastpage}

\end{document}